\documentclass[acmsmall]{acmart}
\AtBeginDocument{%
  }

\newcommand{\revone}[1]{\textcolor{black}{#1}}
\newcommand{\revtwo}[1]{\textcolor{black}{#1}}
\newcommand{\revthr}[1]{\textcolor{black}{#1}}

\usepackage{booktabs}
\usepackage{colortbl}
\usepackage{tablefootnote}
\usepackage[ruled]{algorithm2e}
\usepackage{multirow}
\usepackage{graphicx}
\usepackage{tabularx}
\usepackage{float} 
\usepackage{enumitem}
\usepackage{amsmath, booktabs, threeparttable}
\usepackage{graphicx}
\usepackage{subcaption}

\setcopyright{cc}
\setcctype{by}
\acmJournal{PACMMOD}
\acmYear{2025} \acmVolume{3} \acmNumber{6 (SIGMOD)} \acmArticle{287} \acmMonth{12} \acmPrice{}\acmDOI{10.1145/3769752}

\begin{document}

\title{A General Framework for Per-record Differential Privacy}

\author{Xinghe Chen}
\affiliation{%
  \institution{Nanyang Technological University}
  \city{Singapore}
  \country{Singapore}
}
\email{xinghe001@e.ntu.edu.sg}

\author{Dajun Sun}
\affiliation{%
  \institution{Hong Kong University of Science and Technology}
  \city{Hong Kong}
  \country{China}
}
\email{dsunad@connect.ust.hk}

\author{Quanqing Xu}
\affiliation{%
  \institution{OceanBase, Ant Group}
  \city{Hangzhou}
  \country{China}
}
\email{xuquanqing.xqq@oceanbase.com}

\author{Wei Dong}
\authornote{Wei Dong is the corresponding author.}
\affiliation{%
  \institution{Nanyang Technological University}
  \city{Singapore}
  \country{Singapore}
}
\email{wei_dong@ntu.edu.sg}

\renewcommand{\shortauthors}{Xinghe Chen, Dajun Sun, Quanqing Xu, and Wei Dong}

\begin{abstract}
    Differential Privacy (DP) is a widely adopted standard for privacy-preserving data analysis, but it assumes a uniform privacy budget across all records, limiting its applicability when privacy requirements vary with data values. Per-record Differential Privacy (PrDP) addresses this by defining the privacy budget as a function of each record, offering better alignment with real-world needs.
    However, the dependency between the privacy budget and the data value introduces challenges in protecting the budget’s privacy itself.
    Existing solutions either handle specific privacy functions or adopt relaxed PrDP definitions. A simple workaround is to use the global minimum of the privacy function, but this severely degrades utility, as the minimum is often set extremely low to account for rare records with high privacy needs.
    In this work, we propose a general and practical framework that enables any standard DP mechanism to support PrDP, with error depending only on the minimal privacy requirement among records actually present in the dataset. Since directly revealing this minimum may leak information, we introduce a core technique called \emph{privacy-specified domain partitioning}, which ensures accurate estimation without compromising privacy. We also extend our framework to the local DP setting via a novel technique, \emph{privacy-specified query augmentation}. Using our framework, we present the first PrDP solutions for fundamental tasks such as count, sum, and maximum estimation. Experimental results show that our mechanisms achieve high utility and significantly outperform existing Personalized DP (PDP) methods, which can be viewed as a special case of PrDP with relaxed privacy protection.
\end{abstract}



\received{April 2025}
\received[revised]{July 2025}
\received[accepted]{August 2025}

\maketitle

{
    \section{Introduction}
\label{sec:intro}

\textit{Differential privacy} (DP) is a rigorous standard for protecting individual information in private data release.
It has been widely adopted in the industry~\cite{apple_dp_2017, erlingsson2014rappor, ding2017collecting} or government census authority~\cite{machanavajjhala2008privacy}.
Informally, DP ensures that the presence or absence of a single record in a dataset is $\varepsilon$-indistinguishable from the query results.
The parameter $\varepsilon$, known as the privacy budget, controls the strength of the privacy protection—a smaller value indicates stronger protection.
Although the standard DP model is widely adopted, it assumes uniform privacy requirements across all records. 
This assumption is undesirable in many cases.
For example, when a bank aims to estimate the total deposits of its customers, account records associated with large balances may require stronger privacy protection.  
In this case, each record can have its own privacy budget, which can be formulated as a function $\mathcal{E}(r)$ of the record $r$.
In the above example, where a larger balance requires higher privacy protection, we can define the privacy function to be inversely proportional to the balance. Such a setting is called per-record DP (PrDP)~\cite{alaggan2015heterogeneous, seeman2023privately}. 
Compared with the standard DP model, the PrDP model provides record-dependent privacy protection, thus it is more flexible and can better fit some realistic requirements, like the above example of bank deposit estimation.

Recently, another DP model called personalized DP (PDP)~\cite{ jorgensen2015conservative, dajun2024PDP} has been introduced to meet personalized privacy needs, where each data record is assigned a fixed, publicly known privacy budget. 
In contrast, PrDP employs a privacy budget function $\mathcal{E}(\cdot)$ that adapts the budget based on data values. 
While these models may appear distinct at first glance, we can show that PrDP is a more general notation with stronger privacy guarantees. 
Specifically, by assigning a user-specified attribute to each record and defining the privacy budget function based on this attribute, any PDP problem can be reduced to a PrDP problem. 
Moreover, this reduction offers stronger privacy protection, as the user-defined privacy budgets remain private under PrDP, whereas PDP requires them to be publicly disclosed (see Section~\ref{sec:prdp_pdp} for more details).

Although PrDP is more general than both DP and PDP, it remains challenging to realize in practice. Despite being proposed over a decade ago, no existing work has fully achieved PrDP, even for specific query classes. 
For example, \cite{alaggan2015heterogeneous} studies sum estimation under PrDP with a highly restricted privacy function—proportional to the data value—and shows that the Laplace mechanism satisfies PrDP in this special case. 
Meanwhile, \cite{seeman2023privately} considers more general privacy functions, but under a relaxed version of the PrDP definition. In the following, we identify two key challenges in achieving PrDP.

\begin{table*}[h]
    \centering
    \resizebox{1\columnwidth}{!}{\renewcommand{\arraystretch}{1.5}    
    \begin{tabular}{c c c c}
        \hline
        Task & Basic Counting & Sum Estimation & Distinct Count \\
        \hline\hline

        $\varepsilon$-DP 
        & $O\left(1 / \varepsilon\right)$ 
        & $O\left(\log\log U \cdot\text{Max}(D)  / \varepsilon\right)$~\cite{dong2023universal} 
        & $O\left(1 / \varepsilon\right)$ \\
        \hline

        PDP 
        & $O\left(\log\frac{\hat{\varepsilon}}{\check{\varepsilon}} \log\log\frac{\hat{\varepsilon}}{\check{\varepsilon}}\cdot1 / \varepsilon_{\min}(D)\right)$~\cite{dajun2024PDP}
        & $O\left(\log\frac{\hat{\varepsilon}}{\check{\varepsilon}}  \log\log\frac{\hat{\varepsilon}}{\check{\varepsilon}}\cdot\text{Max}(D) / \varepsilon_{\min}(D)\right)$~\cite{dajun2024PDP}
        & N.A. \\
        \hline

        PrDP (Ours) 
        & $O\left(\log \log\frac{\hat{\varepsilon}}{\check{\varepsilon}}\cdot 1 / \varepsilon_{\min}(D) \right)$ 
        & $O\left((\log \log\frac{\hat{\varepsilon}}{\check{\varepsilon}} + \log \log U)\cdot \text{Max}(D) / \varepsilon_{\min}(D) 
        \right)$ 
        & $O\left(\log \log\frac{\hat{\varepsilon}}{\check{\varepsilon}}\cdot 1 / \varepsilon_{\min}(D)\right)$ \\
        \hline

        $\varepsilon$-LDP 
        & $O\left(\sqrt{n} / \varepsilon\right)$ 
        & $O\left(\sqrt{n}\cdot\log U \cdot \log \log U \cdot \text{Max}(D) / \varepsilon\right)$~\cite{huang21mean}
        & N.A. \\
        \hline

        PrLDP (Ours) 
        & $O\left(\sqrt{n} \cdot \log \log\frac{\hat{\varepsilon}}{\check{\varepsilon}} \cdot 1/ \varepsilon_{\min}(D)\right)$ 
        & $O\left(\sqrt{n} \cdot  (\log \log\frac{\hat{\varepsilon}}{\check{\varepsilon}} +  \log U \cdot \log \log U)\cdot\text{Max}(D) / \varepsilon_{\min}(D)\right)$ 
        & N.A. \\
        \hline
    \end{tabular}
    }
    \vspace{0.1in}
    \caption{Comparison of error bounds under different privacy models across tasks.}   \label{tab:privacy_comparison}
\end{table*}

\textbf{Challenge 1 (Utility).} To achieve PrDP, a naive approach is to use the global minimum privacy budget $\check{\varepsilon}$ for all records, thereby reducing PrDP problems to standard DP problems. 
This leads to a $\check{\varepsilon}$-dependent error bound.
However, as a global lower bound on privacy budgets, $\check{\varepsilon}$ is typically set to a conservatively small value. 
It corresponds to a hypothetical record with the most stringent privacy requirement—an extreme case rarely observed in practice. 
\revtwo{In practice, $\check{\varepsilon}$ is also associated with a global maximum value, denoted by $U$. 
For example, financial institutions often set the maximum bank account deposit to $U = 10^{12}$, equivalent to 1,000 billion~\$. 
If we use the privacy function discussed earlier, then $\check{\varepsilon} \sim 10^{-12}$.}
As a result, using such a $\check{\varepsilon}$ leads to significant utility degradation, as most existing DP mechanisms have an error proportional to $1/\varepsilon$.
Instead, it is preferable that, for a specific dataset $D$, the error depends on the minimal privacy budget within the instance, denoted as $\varepsilon_{\min}(D)$. 
As in the example of bank total deposit estimation, $\varepsilon_{\min}(D)$ depends on the records with the maximum balance in the dataset, which is much smaller than $10^{12}$.
To address this issue, existing research assumes a tighter $\check{\varepsilon}$~\cite{papernot2018scalable,  feldman2021individual, redberg2021privately}, which is close to $\varepsilon_{\min}(D)$. 
However, such prior knowledge compromises privacy, as $\varepsilon_{\min}(D)$ is itself correlated with the records.

\textbf{Challenge 2 (Privacy).} 
To avoid relying on the global minimum privacy budget, a common strategy is the \textit{clipping mechanism}~\cite{abadi2016deep, huang21mean}. 
For example, in deposit estimation, all values above a threshold $\tau$ are clipped to $\tau$, and a DP sum estimator is applied to the clipped data using a privacy budget $\mathcal{E}(\tau)$.
However, this approach further presents two problems. 
First, utility highly depends on the choice of $\tau$, which ideally should be dataset-dependent. 
For example, to achieve an error dependency on $\varepsilon_{\min}(D)$, we can set $\tau$ to the maximum deposit in $D$, $\text{Max}(D)$—a sensitive value that must itself be estimated under PrDP—thus reintroducing the need to select an appropriate privacy budget.
Besides choosing an appropriate $\tau$, another problem is that such an idea will break the privacy of the clipped records.
This is because the privacy budget is determined by the original values and is unaffected by clipping.  
Specifically, even if the balance of $r$ is clipped to $\tau$, its privacy budget must still be $\mathcal{E}(r)$ rather than $\mathcal{E}(\tau)$.  
To address this, Jeremy Seeman et al.~\cite{seeman2023privately} propose an alternative that sacrifices privacy guarantees for records exceeding $\tau$. 
However, this approach relies heavily on an appropriate choice of $\tau$, which requires strong prior knowledge, and fails to fully satisfy the requirements of PrDP.

In addition to these two challenges, PrDP solutions should also be general, i.e., not only supporting arbitrary privacy budgets, but also accommodating a wide range of query types. Above all, this raises a question:

\textbf{Problem Statement.} Does there exist a general framework that enables an arbitrary DP mechanism $\mathcal{M}$ to provide a PrDP guarantee with error dependency on $\varepsilon_{\min}(D)$?

\subsection{Our Results}  
\label{sec:intro_results}
In this paper, we answer the question affirmatively by presenting a novel general framework that enables existing DP mechanisms to solve PrDP problems. 
\revtwo{To address the issue of the clipping mechanism, we first propose a technique called \textit{privacy-specified domain partitioning}, which partitions records based on their privacy budgets.} 
This method partitions the entire record universe into $\lceil \log (\hat{\varepsilon} / \check{\varepsilon}) \rceil$ disjoint privacy-specified domains, where $\hat{\varepsilon}$ denotes the predefined global maximum privacy budget. 
Within each domain, all records have similar privacy budgets.
We then estimate the number of records in $D$ for each domain using the smallest privacy budget assigned to that domain. 
Based on these estimates, we identify the ``heavy'' domains and use the smallest privacy budget among them, denoted by $\varepsilon_\tau$, to approximate $\varepsilon_{\min}(D)$.
Naturally, this idea leads directly to a PrDP counting algorithm by aggregating the counting of ``heavy'' domains.
For other types of queries, we further take the union of all domains with privacy budgets exceeding $\varepsilon_\tau$, and construct a dataset containing only the records within this unioned domain. Finally, we invoke an existing DP mechanism on the new dataset with privacy budget $\varepsilon_\tau$ to compute the final query results.
Building on this idea, we can extend any standard DP protocol to satisfy PrDP, with an error that depends on $\varepsilon_{\min}(D)$.

Our main contributions are threefold:

\begin{itemize}
    \item \textbf{Basic counting under PrDP.} Section~\ref{sec:first_attempt} introduces our approach to the basic counting problem under PrDP. 
    Algorithm~\ref{alg:cnt} achieves an error bound of $O\left(\log \log(\hat{\varepsilon}/\check{\varepsilon}) / \varepsilon_{\min}(D)\right)$\footnote{All error bounds stated in the introduction holds with constant probability.}. 
    Compared to the standard DP error bound of $O(1/\varepsilon)$, our mechanism replaces $\varepsilon$ with $\varepsilon_{\min}(D)$, incurring only a doubly logarithmic overhead.

    \item \textbf{General problems under PrDP.} Section~\ref{sec:gen_fmk_PrDP} presents a general framework for arbitary PrDP problems, built upon our basic counting algorithm. 
    The error bound of this framework is somewhat technical.
    Roughly speaking, given any standard DP mechanism $\mathcal{M}$ with an error $\text{Err}_{\mathcal{M}}(D, \, \varepsilon)$ on $D$, our framework almost achieves an error $\tilde{O}(\text{Err}_{\mathcal{M}}(D, \, \varepsilon_{\min}(D)))$~\footnote{We use $\tilde{O}(\cdot)$ to hide polylogarithmic factors.} (see Theorem~\ref{the:fmk_bound} for more details).

    \item \textbf{Problems under local-DP setting.}
    \revtwo{In Section~\ref{sec:PrLDP}, we extend our PrDP framework to the local DP (LDP) setting—speci-fically, per-record local DP (PrLDP)—by introducing a novel technique called \textit{privacy-specified query augmentation}}.
    Our framework achieves results comparable to those in the central DP setting. 
    Given an LDP protocol $\mathcal{M}$ with error $\text{Err}_{\mathcal{M}}(D, \,$ 
    $\varepsilon)$ on dataset $D$, our PrLDP framework achieves an error of approximately $\tilde{O}(\text{Err}_{\mathcal{M}}(D, \, \varepsilon_{\min}(D)))$ (see Theorem~\ref{the:ldp_fmk_bound} for details).
\end{itemize}

For error bounds on specific queries within our general framework, refer to Table~\ref{tab:privacy_comparison}.
To the best of our knowledge, this is the first work to achieve PrDP even for fundamental queries. As a baseline, we compare our results with those from~\cite{dajun2024PDP}, the state-of-the-art (SOTA) PDP mechanisms for counting and sum estimation. 
This comparison is not entirely fair since, as previously noted, PDP is a special case of PrDP with a relaxed privacy guarantee. 
Surprisingly, we even improved their PDP error bounds, reducing the polylogarithmic factor to a doubly logarithmic one.
Moreover, in Section~\ref{sec:experiment}, we demonstrate our framework yields  $2\times$ to $165\times$ reduction in error compared to prior PDP methods in practice~\footnote{Code is available at https://github.com/XChen1998/A-General-Framework-for-Per-record-Differential-Privacy.}. 
\revone{Its generalizability is also verified under three different privacy budget functions: inverse, logarithmic, and square root.}

    \section{Related Work}
\label{sec:re_work}
Providing record-specific privacy protection is a highly relevant and important problem, driven by the diverse backgrounds of users and their varying privacy requirements. 
Such a challenge was explored even before the concept of DP. 
For instance, Xiao and Tao~\cite{xiao2006personalized} introduced the concept of personalized privacy within the framework of $k$-anonymity, which was subsequently extended to other privacy models~\cite{ye2008personalized, gedik2007protecting, shen2009research}. 

Under the context of DP~\cite{cynthia2009roboust, dwork2006privacyBook}, which provides a rigorous definition of privacy and has become the dominant privacy model, several approaches offer record-specific privacy protection.
PrDP was introduced by \cite{alaggan2015heterogeneous} and formalized in \cite{seeman2023privately}. 
They studied some basic queries under specific privacy functions. 
However, they do not support arbitrary privacy functions, even for their selected queries.
\revtwo{Besides PrDP, Jorgensen \textit{et al.}~\cite{jorgensen2015conservative} propose the PDP model, where each user specifies a (public) personalized privacy parameter.}
As we show in Section~\ref{sec:prdp_pdp}, this can be regarded as a special case of PrDP with relaxed privacy protection.
Significant efforts have been made to achieve PDP under various settings~\cite{ebadi2015differential, li2017partitioning, cui2019improving, niu2021adapdp, redberg2021privately, boenisch2023have, yang2023personalized, liu2024cross}.  
Most recently, Sun \textit{et al.}~\cite{dajun2024PDP} provided the first formal error bounds for PDP on fundamental queries, including basic counting and sum estimation.

Moving towards the LDP setting, existing studies have primarily focused on PDP~\cite{chen2016private, yiwen2018utility, xue2022mean, bao2021successive, li2022protecting}, while no prior work has addressed PrDP. In contrast, our work is the first to investigate PrDP in the local setting.

    \section{Preliminary}
\label{sec:prelim}

\subsection{Notation}

Let $[U]=\{0,\,1,\,\ldots,\,U\}$.
A dataset $D = \{r_1, \, r_2, \, \ldots, \, r_n\}$ contains $n \in \mathbb{N}$ records, where each record $r \in [U]^d$ is a $d$-dimensional vector. 
We say that $D' \subseteq D$ if $D' = D$ or if $D'$ can be obtained by deleting some records from $D$.  
Let $|\cdot|$ denote the $\ell_1$ norm for any vector.
As in the bank example provided in the introduction, the dataset $D$ contains the account information of a bank's customers, with each record $r$ represented as
\begin{equation*}
r = (v_\text{Bal}, \, v_\text{PC}, \, \ldots).
\end{equation*}
where each coordinate represents a different attribute of the customer such as account balance ($v_\text{Bal}$) and mailing postcode ($v_\text{PC}$).
Note that the integer domain is a standard and widely adopted setting in the DP literature~\cite{dong2023universal, dong2022r2t, huang21mean}, as fractional domains can be easily mapped to integers.  
For example, a bank account balance, typically represented as a decimal number, can be converted to an integer by defining a base unit and applying quantization.

\subsection{DP Basics}

In this paper, we define two datasets, $D$ and $D'$, as neighbors if they differ by the insertion or deletion of a single record, denoted as $D \sim D'$.  
We use $D \sim_r D'$ to indicate that $D$ and $D'$ differ specifically on the record $r$.
We define $d(D,\, D')$ as the distance between $D$ and $D'$, measured by the minimum number of single-record additions or deletions needed to transform $D$ into $D'$.
The standard DP model is defined as follows:  

\begin{definition}
    \emph{(Differential Privacy).} For $\varepsilon > 0$, a mechanism $\mathcal{M}$ is $\varepsilon$-DP if for any pair of neighboring datasets $D \sim_r D'$ and any possible output $y$, the following holds:
    \begin{equation*}
        \Pr[\mathcal{M}(D) = y] \leq e^\varepsilon \cdot \Pr[\mathcal{M}(D') = y].
    \end{equation*}
\end{definition}

\subsubsection{Laplace mechanism}

A commonly used one-dimensional DP mechanism is the \textit{Laplace mechanism}. 
Given any query $Q : [U]^{n \times d} \to \mathbb{R}$, its \textit{global sensitivity} ($\text{GS}_Q$) is defined as  
\begin{equation*}
    \text{GS}_Q = \max_{\substack{D, \, D' \, : \, D \sim D'}} |Q(D) - Q(D')|.
\end{equation*}
The Laplace mechanism is given by

    


\begin{lemma}
\label{lem:Lap}
    \emph{(Laplace Mechanism)}. Given a query $Q : [U]^{n\times d} \to \mathbb{R}$, the Laplace mechanism  
    \begin{equation*}
        \mathcal{M}(D) = Q(D) + \eta
    \end{equation*}
    satisfies $\varepsilon$-DP, where $\eta$ is drawn from a Laplace distribution with scale $\text{GS}_Q/\varepsilon$, i.e., $\eta \sim \text{Lap} \left( \text{GS}_Q/\varepsilon \right)$.
\end{lemma}

The following Laplace tail bound shows that the error of the Laplace mechanism is at most $\text{GS}_Q/\varepsilon \cdot \log(1/\beta)$ with probability at least $1 - \beta$.  

\begin{lemma}
\label{lem:tail}
    \emph{(Laplace Tail Bound)}. Let $\eta \sim \text{Lap}(1)$ be a random variable drawn from the unit Laplace distribution.  
    For any $\beta \in (0, 1)$, the tail probability of $\eta$ satisfies  
    \begin{equation*}
        \Pr(|\eta| \geq \ln\frac{1}{\beta}) \leq \beta.
    \end{equation*}
\end{lemma}

Therefore, for a basic counting problem under $\varepsilon$-DP, where $\text{GS}_Q = 1$, the Laplace mechanism achieves an error of $O\left(1/\varepsilon\right)$ with probability at least $1 - \beta$.
However, for more complicated problems like one-dimensional sum estimation, where $\text{GS}_Q = U$, this mechanism incurs an error of $O(U/\varepsilon)$. 
This is undesirable, as $U$ is a predefined global parameter that must be set conservatively large to accommodate all possible datasets.  
For example, in the bank deposit estimation problem, $U$ is set to $10^{12}$ to handle the largest possible deposit, but such a value rarely appears in most banks.

In contrast, recent studies have achieved an instance-specific error bound of $\tilde{O}\left(\max(D)/\varepsilon\right)$~\cite{dong2024instance}, where $\max(D)$ is the largest value in $D$ and $\max(D) \ll U$ for most datasets.  
In the bank total deposit problem, $\text{Max}(D) = \max\{v_{\text{Bal}} \mid r \in D\}$ represents the highest balance in $D$ considered in this study, which is significantly smaller than $10^{12}$.  

\subsubsection{Optimality of DP Errors}  
\label{sec:opt_DP}  

The Laplace mechanism is known to achieve \textit{worst-case optimal error}~\cite{dong2024almost}.  
However, worst-case optimality often introduces an unnecessarily large error in most instances.  
To mitigate this, \textit{down-neighborhood optimality}~\cite{fang2022shifted} was introduced as a more meaningful criterion for defining optimality in standard DP error.  

\begin{definition}
\emph{(Down-neighborhood Optimality)}~\cite{fang2022shifted}.  
Let $\mathbb{M}$ be the class of all $\varepsilon$-DP mechanisms.  
Given any query $Q$, the $\rho$-down neighborhood lower bound on dataset $D$ is defined as  
\begin{align*}
    & \mathcal{L}(D, \rho) :=  \\
    & \quad\inf_{\mathcal{M'} \in \mathbb{M}} \sup_{\substack{D' \subseteq D, \\ d(D,\, D') \leq \rho}} \inf \Big\{ \xi :  
    \Pr \big[ |\mathcal{M'}(D') - Q(D')| \leq \xi \big] \geq \frac{2}{3} \Big\},
\end{align*}
and a mechanism $\mathcal{M}$ is $(\rho, c)$-down neighborhood optimal if, for any dataset $D$,  
\begin{equation*}
    \Pr \left[ |\mathcal{M}(D) - Q(D)| \leq c \cdot \mathcal{L}(D, \, \rho) \right] \geq \frac{2}{3},
\end{equation*}
where $c$ is a constant quantifying the level of optimality.  
\end{definition}

Roughly speaking, a mechanism $\mathcal{M}$ is $(\rho,c)$-down neighborhood optimal if, for every $D$, it performs as well as the optimal mechanism $\mathcal{M}'$ specifically designed for $D$ and its $\rho$-down neighborhood.  
Clearly, smaller $\rho$ and $c$ imply stronger optimality.
Moreover, $\mathcal{L}(G, \rho)$ is further lower-bounded by the \textit{$(\rho-1)$-downward query difference}.

\begin{theorem}[\cite{fang2022shifted}]
\label{the:dn_bound}
For any query $Q$, dataset $D$, $\varepsilon \leq \ln 2$ and $\rho \geq 1$, we have  
\begin{equation*}
    \mathcal{L}(D, \rho) \geq \frac{\Delta Q^{(\rho-1)}(D)}{2\rho},
\end{equation*}
where  
\begin{equation*}
    \Delta Q^{(\rho-1)}(D) = \max_{\substack{D' \subseteq D,\,  d(D, D') \leq \rho}} \big|Q(D') - Q(D)\big|
\end{equation*}
is the \textit{$(\rho-1)$-downward query difference}.
\end{theorem}

Therefore, if a DP mechanism achieves an error of $c\cdot\Delta Q^{(\rho-1)}(D)$, then it is $(\rho,\, 2c\rho)$-down neighborhood optimal.  
It is widely accepted that when $\rho = \tilde{O}(1/\varepsilon)$, such an error is good enough~\cite{fang2022shifted, dong2023universal}.  
Moreover, to the best of our knowledge, any existing DP mechanism incurs an error of $\Omega(\Delta Q^{\rho}(D))$ with $\rho = \Omega(1/\varepsilon)$.



\subsection{Per-record Differential Privacy}
\label{sec:prdp_prelim}
The standard DP model assumes that all records have equal privacy requirements.  
However, in reality, privacy concerns may vary across records depending on their content, as discussed in the introduction.

To accommodate diverse real-world needs, the per-record DP framework~\cite{alaggan2015heterogeneous, seeman2023privately} was introduced, allowing each record to have a privacy requirement based on its content.  
In this framework, a record-dependent privacy budget function $\mathcal{E}(\cdot)$ is provided to define the privacy requirement of each record $r$.

\begin{definition}
\label{def:prpbfunc}
    \emph{(Record-Dependent Privacy Budget Function)}.  
    A record-dependent privacy budget function, or simply a privacy budget function, is a mapping $\mathcal{E}: [U]^d \to [\check{\varepsilon}, \hat{\varepsilon}]$,  
    where $\check{\varepsilon}, \hat{\varepsilon} \in \mathbb{R}_{+}$ are predefined lower and upper bounds on the privacy budgets, respectively.
\end{definition}

For the example of estimating a bank's total deposits, a possible privacy budget function is defined as  
\begin{equation}
\label{eqn:pbf_exp}
    \mathcal{E}(r) =
    \begin{cases}
        \alpha / v_{\text{Bal}}, & \text{if } \alpha / \hat{\varepsilon} \leq v_{\text{Bal}}, \\
        \hat{\varepsilon}, & \text{if } v_{\text{Bal}} > \alpha / \hat{\varepsilon}
    \end{cases}
\end{equation}
where $\alpha$ is a constant factor and $\hat{\varepsilon}$ is a predefined upper bound  on the privacy budget.
Since $v_{\text{Bal}}$ has a domain of $[U]$, there is no need to explicitly specify $\check{\varepsilon}$; instead, we use $\check{\varepsilon}=\alpha/U$. 
This privacy budget function ensures stronger protection for records with larger values.

%

With the privacy budget function, we can define per-recond DP:



\begin{definition}
    \emph{(Per-record Privacy (PrDP))}. Given a record-dependent privacy budget function $\mathcal{E}:[U]^d \to [\check{\varepsilon},\, \hat{\varepsilon}]$, a mechanism $\mathcal{M}$ is  $\mathcal{E}$-per-record differentially private, or simply $\mathcal{E}$-PrDP, if for any pair of neighboring datasets $D \sim_r D'$ and for all possible outputs $y$, the following holds:
    \begin{equation*}
        \Pr[\mathcal{M}(D) = y] \leq e^{\mathcal{E}(r)} \cdot \Pr[\mathcal{M}(D') = y].
    \end{equation*}
\end{definition}




As required by our methodology, we define the concept of the \textit{privacy-specified domain} as follows:  

\begin{definition}
\label{def:prdmn}
    \emph{(Privacy-Specified Domain)}. For a given $\mathcal{E} : [U]^{d} \to \mathbb{R}_+$ and a privacy budget range $[\varepsilon_{a}, \, \varepsilon_{b}]$, the privacy-specified domain is defined as  
    \begin{equation*}
        \mathcal{X}_{[\varepsilon_{a}, \, \varepsilon_{b}]} = \{r\in [U]^d  \mid \mathcal{E}(r) \in [\varepsilon_{a}, \, \varepsilon_{b}]\}.
    \end{equation*}
\end{definition}

For example, if the privacy budget function is given by (\ref{eqn:pbf_exp}) with $\alpha=10^4$, then $\mathcal{X}_{[0.002, \, 0.004]}$ includes records with $v_{\text{Bal}}\in [2.5\times10^6,\, 5\times10^6]$.  

PrDP enjoys the following general properties:

\begin{lemma}
\label{lem:post}
    \emph{(Post-Processing)}. If a mechanism $\mathcal{M}$ is considered $\mathcal{E}$-PrDP, then for any mechanism $\mathcal{M}'$, $\mathcal{M}'(\mathcal{M}(D))$ is still $\mathcal{E}$-PrDP.
\end{lemma}

\begin{lemma}
\label{lem:seq}
    \emph{(PrDP Sequential Composition)}. Let $\{\mathcal{M}_1,\, \ldots,\, \mathcal{M}_k\}$ be a series of $\mathcal{E}_i$-PrDP mechanisms. Then $\mathcal{M}_{[k]}$, defined as  
    \begin{equation*}
        \mathcal{M}_{[k]}(D) := (\mathcal{M}_1(D),\, \mathcal{M}_2(D),\, \ldots,\, \mathcal{M}_k(D)),
    \end{equation*}   
    satisfies $\mathcal{E}'$-PrDP, where $\mathcal{E}'(r) = \sum_{i=1}^{k} \mathcal{E}_i(r)$ for all $r \in [U]^d$. 
\end{lemma}

\begin{lemma}
\label{lem:para}
    \emph{(PrDP Parallel Composition)}. Let $\mathcal{R}_1, \dots, \mathcal{R}_k$ be a set of disjoint subdomains.  
    Let $\{\mathcal{M}_1, \dots, \mathcal{M}_k\}$ be a series of $\mathcal{E}_i$-PrDP mechanisms.  
    Then their parallel composition  
    \begin{equation*}
        \mathcal{M}(D) := \left(\mathcal{M}_1(D \cap \mathcal{R}_1), \dots, \mathcal{M}_k(D \cap \mathcal{R}_k) \right)
    \end{equation*}
    satisfies $\mathcal{E}'$-PrDP, where $\mathcal{E}'(r) = \max\{\mathcal{E}_i(r)\}_{i=1}^k$ for all $r \in [U]^d$.
\end{lemma}



\subsection{PrDP and PDP}
\label{sec:prdp_pdp}

PDP assumes that each record $r_i$ has a user-defined privacy budget $\varepsilon_i$ that is independent of its value.  
In contrast, PrDP defines the privacy budget of $r_i$ as a function of its value.  
Furthermore, PDP considers all record values $r_i$ as public information, whereas PrDP only reveals the privacy budget function $\mathcal{E}(\cdot)$ while keeping the privacy budget of each individual record private.

At first glance, both PDP and PrDP provide record-specific privacy protection but follow different models and are thus incomparable.  
PDP defines the privacy protection level based on user-specified budgets, whereas PrDP determines it as a function of each record’s value.  
In the following example, we show that PDP is, in fact, just a special case of PrDP with a relaxed privacy guarantee.


Recall that each record $r_i$ is a high-dimensional vector. We introduce an additional user-specified attribute, $v_{\text{bud}}$, and define the privacy budget function as $\mathcal{E}(r) = c \cdot v_{\text{bud}}$, where $c$ is a predefined parameter. 
This design enables a user-specific privacy budget for each record, achieving the functionality of PDP. 
Furthermore, in this setting, PrDP maintains the privacy of each record's budget, whereas PDP requires it to be publicly disclosed.





    \section{Straw-man Framework}
\label{sec:straw}

We begin by exploring several straightforward approaches to achieve $\mathcal{E}$-PrDP.  
While these methods ultimately fail to achieve $\mathcal{E}$-PrDP, they highlight key challenges and offer valuable insights for our final solution.


\subsection{Utilize the Global Minimal Privacy Budget}
\label{sec:min_privacy_budget}

The most intuitive way to satisfy $\mathcal{E}$-PrDP is to utilize the global minimal privacy budget $\check{\varepsilon}$, thereby reducing the problem to $\check{\varepsilon}$-DP.  
In the example of bank deposit estimation, the smallest possible privacy budget is $\check{\varepsilon} = \alpha / U$.  
The best error for one-dimensional sum estimation under standard DP is $O\left(\log\log U \cdot\text{Max}(D)/\varepsilon  \right)$~\cite{dong2023universal}.  
Substituting $\check{\varepsilon} = \alpha / U$ yields an error of $\tilde{O}(\text{Max}(D)\cdot U)$.  
Clearly, using $\check{\varepsilon}$ directly leads to excessively large errors, as demonstrated in the introduction.  
This occurs because the approach always considers the record with the strongest global protection, even if such records may not appear in most datasets.  

Therefore, achieving an error dependency on the instance-specific minimal privacy budget, i.e., $\varepsilon_{\min}(D)$, is much more preferable.  
In the previous example, using $\varepsilon_{\min}(D)$ as the privacy budget results in an error of $\tilde{O}\left(\text{Max}(D)^2\right)$, which can be significantly smaller since $ \text{Max}(D) \ll U$.  
However, directly utilizing $\varepsilon_{\min}(D)$ violates PrDP, since it depends on $\mathrm{Max}(D)$, a highly sensitive statistic. This motivates our development of alternative solutions.




\subsection{Clipping Mechanism in PrDP}
\label{sec:clip_prdp}

A natural idea to reduce the error dependence on the global minimum privacy budget $\check{\varepsilon}$ to an instance-specific budget, $\varepsilon_{\min}(D)$, is the clipping mechanism.  
Such a technique is widely used in standard DP sum estimation to reduce error dependence from a predefined domain upper bound $U$ to $\text{Max}(D)$.  
The high-level idea is to first estimate a threshold $\tau \sim \text{Max}(D)$ and then clip values over $\tau$ to $\tau$.  
By adding noise proportional to $\tau$ to the result over the clipped data, we can achieve an error depending on $\text{Max}(D)$.  

Inspired by the clipping mechanism under standard DP, except for finding a value threshold $\tau$, our idea is to also privately find a suitable estimation of $\varepsilon_{\min}(D)$, denoted as $\varepsilon_\tau$.  
Then, we clip the records with a privacy requirement stronger than $\varepsilon_\tau$ and run an $\varepsilon_\tau$-DP mechanism on the clipped dataset.  
In the example of bank deposit estimation, since the privacy budget is inversely proportional to $v_{\text{Bal}}$, estimating $\varepsilon_\tau$ is equivalent to estimating $\text{Max}(D)$.  
Thus, we first find a close approximation of $\text{Max}(D)$, denoted as $\tau$, then clip all records beyond this threshold.  
Finally, we invoke an $\varepsilon_{\tau}$-DP mechanism over the clipped data with $\varepsilon_{\tau} = \mathcal{E}(\tau)$.  

At first glance, this method appears to achieve $\mathcal{E}$-PrDP, as clipping ensures that all records are bounded by $\tau$, with the privacy budgets bounded by $\varepsilon_{\tau}$.  
However, we will show below that this approach neither satisfies PrDP nor guarantees an error dependency on $\varepsilon_{\min}(D)$.
\begin{itemize}  
    \item \textbf{Challenge 1 (Privacy)}:  
    \emph{Invoking a DP mechanism with $\varepsilon_\tau$ over clipped data is insufficient to guarantee $\mathcal{E}$-PrDP.}   
    By definition, the privacy budget of a record $r$ should be $\mathcal{E}(r)$. 
    Thus, even if we clip its value to $\tau$, we must still adhere to its original privacy budget rather than the privacy budget after clipping. 
    Therefore, $\varepsilon_\tau$ is insufficient to guarantee privacy.
    
    \item \textbf{Challenge 2 (Utility)}:  
    \emph{Accurately estimating $\varepsilon_{\min}(D)$ under $\mathcal{E}$-PrDP is challenging.}
    To estimate $\varepsilon_{\min}(D)$ under $\mathcal{E}$-PrDP, the foremost issue is determining the appropriate privacy budget to use.
    The most straightforward approach is to use $\check{\varepsilon}$, which leads to an error dependency on $\check{\varepsilon}$. 
    Alternatively, using $\varepsilon_\tau$ to estimate itself results in a circular dependency, creating a chicken-and-egg problem.
\end{itemize}

\section{Our First Attempt in PrDP Protocol Design: Basic Counting}
\label{sec:first_attempt}

In this section, we present our first attempt at designing an $\mathcal{E}$-PrDP protocol by studying the basic counting problem.
This study provides key insights into our design and serves as a foundational building block for our general PrDP framework.
The standard DP protocol for bit counting achieves an error of $O(1/\varepsilon)$ with the Laplace mechanism.
Moving toward $\mathcal{E}$-PrDP, utilizing the global minimum privacy budget $\check{\varepsilon}$ yields an error $O(1/\check{\varepsilon})$.  
Our proposed protocol improves this by reducing the error to $\tilde{O}(1/\varepsilon_{\min}(D))$.  

\subsection{Privacy-Specified Domain Partitioning}
\label{sec:bcnt}

Fundamentally, the first challenge in applying the clipping mechanism to achieve $\mathcal{E}$-PrDP arises from the fact that records with strong privacy requirements retain their original privacy constraints even after clipping. 
Therefore, a natural idea is to directly exclude those records with small privacy budgets. 

This can be achieved through privacy-specified domain partitioning, where the entire domain is divided into sub-domains such that values $r$ within each sub-domain have similar privacy budgets $\mathcal{E}(r)$.  
When we perform count estimation, given a threshold $\varepsilon_{\tau}$, we only add up the counting results for those sub-domains with privacy budgets greater than or equal to $\varepsilon_{\tau}$.  
This approach ensures that records with small privacy budgets (i.e., less than $\varepsilon_{\tau}$) do not incur any privacy leakage, thereby addressing the first challenge associated with the clipping mechanism.

To address the second challenge—accurately estimating $\varepsilon_{\min}(D)$ under PrDP—we extend our domain partitioning approach to identify a threshold $\varepsilon_{\tau}$.  
The goal is to ensure that the count derived from selected sub-domains (i.e., those with privacy budgets greater than $\varepsilon_{\tau}$) can serve as a reliable estimate for the final counting result.  
Since the error consists of two components—the number of records excluded and the noise introduced in the selected sub-domains—achieving an error of $\widetilde{Q}(1/\varepsilon_{\min}(D))$ requires two objectives:

\begin{enumerate}[label=(\alph*)]
    \item At most $\tilde{O}(1/\varepsilon_{\min}(D))$ records are excluded.
    \item The total noise introduced in the counting results of the selected sub-domains stays within $\tilde{O}(1/\varepsilon_{\min}(D))$. 
\end{enumerate}

With these ideas, we now proceed with the design of our protocol.  
For domain partitioning, we begin by segmenting the privacy budget range into $\lceil \log (\hat{\varepsilon} / \check{\varepsilon}) \rceil$ disjoint sub-intervals, where all sub-intervals are left-open and right-closed except for the first one:  
\begin{equation*}
    [2^0 \cdot \check{\varepsilon},\, 2^1 \cdot \check{\varepsilon}], \, 
    (2^1 \cdot \check{\varepsilon},\, 2^2 \cdot \check{\varepsilon}], \,  
    \ldots, \, 
    (2^{\lceil \log (\hat{\varepsilon} / \check{\varepsilon}) \rceil - 1} 
    \cdot \check{\varepsilon},\, \hat{\varepsilon}].
\end{equation*}

By leveraging the concept of a privacy-specified domain, as introduced in Definition~\ref{def:prdmn}, we can partition the input domain into $\lceil \log (\hat{\varepsilon} / \check{\varepsilon}) \rceil$ disjoint sub-domains:  
\begin{equation*}
    \mathcal{X}_{[2^0 \cdot \check{\varepsilon},\, 2^1 \cdot \check{\varepsilon}]}, \, 
    \mathcal{X}_{(2^1 \cdot \check{\varepsilon},\, 2^2 \cdot \check{\varepsilon}]}, \,  
    \ldots, \, 
    \mathcal{X}_{(2^{\lceil \log (\hat{\varepsilon} / \check{\varepsilon}) \rceil - 1} 
    \cdot \check{\varepsilon},\, \hat{\varepsilon}]}.
\end{equation*}

For simplicity, we denote the $i$-th domain $\mathcal{X}_{(2^{i-1} \cdot \check{\varepsilon},\, 2^i \cdot \check{\varepsilon}]}$ as $\mathcal{X}_i$.  
The primary advantage of this partitioning is that for each sub-domain $\mathcal{X}_i$, we can use $2^{i-1} \cdot \check{\varepsilon}$ for all records within that domain.  
Such a design upgrades the privacy protection for each record by at most a factor of two, thereby avoiding the use of an excessively small privacy budget like $\check{\varepsilon}$ for all records.

Then, for each sub-domain $\mathcal{X}_i$, we estimate its counting result with  
\begin{equation}
    \widetilde{Q}_{\text{count}}(D \cap \mathcal{X}_i) := Q_{\text{count}}(D \cap \mathcal{X}_i) + \text{Lap}\left(\frac{1}{2^{i-1} \cdot \check{\varepsilon}} \right).
\end{equation}  
Next, we aim to identify the $\ell$-th domain $\mathcal{X}_\ell$ with its minimum privacy budget $2^{\ell-1} \cdot \check{\varepsilon}$, setting it as $\varepsilon_\tau$, such that $\varepsilon_\tau$ is very close to $\varepsilon_{\min}(D)$.  
Then, we discard all domains with a privacy budget below $\varepsilon_\tau$ and sum up all the noisy counts $\widetilde{Q}_{\text{count}}$ for the remaining domains.
To achieve our targets (a) and (b), $\varepsilon_{\tau}$ should satisfy the following two properties:  
\begin{enumerate}[label=(\roman*)]
    \item $\left|D \cap \left(\bigcup_{i=1}^{\ell-1} \mathcal{X}_i\right) \right| \leq \tilde{O}(1/\varepsilon_{\min}(D))$
    \item $\varepsilon_\tau \geq \frac{1}{2} \varepsilon_{\min}(D)$ 
\end{enumerate}
Property (i) is to ensure objective (a), that no more than $\tilde{O}(1/\varepsilon_{\min}(D))$ records are excluded.  
Property (ii) satisfies (b), ensuring that our estimated $\varepsilon_\tau$ is not overly small, thereby preventing excessive noise.

To achieve these two sub-goals, we identify the first domain $\mathcal{X}_{\ell}$ such that it contains enough records. 
More precisely, from $i \in \{1, \, 2, \, \dots, \, \lceil \log (\hat{\varepsilon} / \check{\varepsilon}) \rceil\}$, we find the first $i$ satisfying
\begin{equation*}
    \widetilde{Q}_{\text{count}}(D \cap \mathcal{X}_i)\geq \frac{1}{2^{i-1} \cdot \check{\varepsilon}} \ln\left(\frac{\lceil \log (\hat{\varepsilon} / \check{\varepsilon}) \rceil}{\beta}\right),
\end{equation*}
where $\beta$ is a predefined failure rate.

The right-hand side serves as the threshold for determining whether there are enough records in $\mathcal{X}_{i}$.
\revtwo{These thresholds are calibrated based on Lemma~\ref{lem:Lap} to satisfy the desired error bound, which is formalized later in Theorem~\ref{the:cnt}.}
This prevents an ``early stop'', where the process halts at an empty $\mathcal{X}_{i}$, thereby satisfying property (ii).
At the same time, the ``late stop'' is also avoided. 
All sub-domains we passed should be very ``light'' meaning there are very few records. 
Otherwise they would pass the thresholds. 
Through a careful calculation, we can show that all passed domains have at most $\tilde{O}(1)$ records thus satisfying property (i).

To summarize, our final protocol works as follows: 
First, it identifies the first $\mathcal{X}_{\ell}$ that contains sufficient records by comparing the query results with carefully calibrated thresholds.
Then, we sum all sub-query results starting from the $\ell$-th privacy-specified domain. 
The detailed algorithm is presented in Algorithm~\ref{alg:cnt}. 
The runtime complexity of this algorithm is $O(n)$.
Our algorithm also returns $\varepsilon_\tau = 2^{\ell-1} \cdot \check{\varepsilon}$ as an estimate of $\varepsilon_{\min}(D)$, which will be further utilized in our general PrDP framework.

\begin{figure*}[t]
    \centering
    \includegraphics[width=\textwidth]{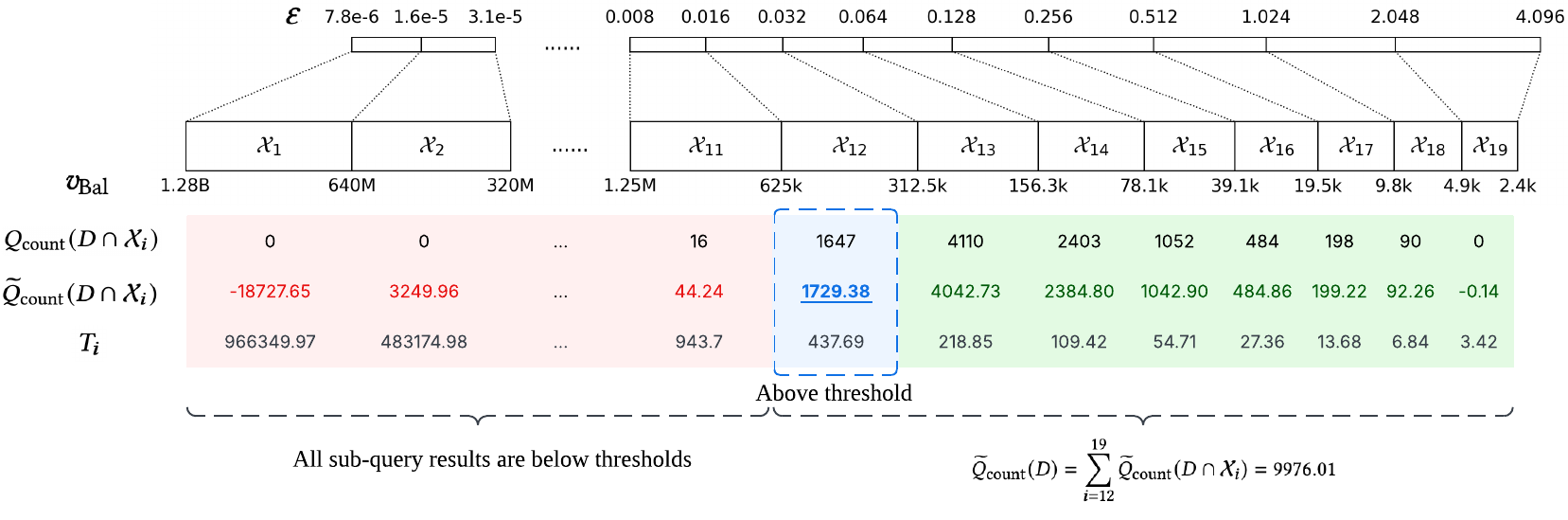}
    \caption{An illustration of Algorithm~\ref{alg:cnt} for a counting problem with the privacy budget function $\mathcal{E}(r) = \alpha / v_{\text{bal}}$, where $\alpha = 10^4$, $U=1,280,000,000$~\$, and $\hat{\varepsilon} = 4.096$.}
    \label{fig:alg_cnt}
\end{figure*}

\begin{algorithm}[t]
\LinesNumbered
\caption{PrDP Count}
\label{alg:cnt}
\KwIn{Privacy budget function $\mathcal{E}(\cdot)$,\, dataset $D$,\, failure rate $\beta$}
\KwOut{$\widetilde{Q}_{\text{count}}(D)$ under $\mathcal{E}$-PrDP}

$\ell \gets \left\lceil \log \frac{\hat{\varepsilon}}{\check{\varepsilon}} \right\rceil$

\tcp*[l]{Obtain results}
\For{$i \gets \{1,\, 2,\, \ldots,\, \lceil \log (\hat{\varepsilon}/\check{\varepsilon}) \rceil\}$}{
    $\widetilde{Q}_{\text{count}}(D \cap \mathcal{X}_i) \gets Q_{\text{count}}(D \cap \mathcal{X}_i) + \text{Lap}\left(\frac{1}{2^{i-1} \cdot \check{\varepsilon}}\right)$
}

\tcp*[l]{Estimate $\ell$, i.e., $\varepsilon_\tau$}
\For{$i \gets \{1,\, 2,\, \ldots,\, \lceil \log (\hat{\varepsilon}/\check{\varepsilon}) \rceil\}$}{
    $\mathcal{X}_i \gets \mathcal{X}_{\left(2^{i-1} \cdot \check{\varepsilon},\, \min\left(2^{i} \cdot \check{\varepsilon},\, \hat{\varepsilon}\right)\right]}$
    
    $T_i \gets \frac{1}{2^{i-1} \cdot \check{\varepsilon}} \ln\left(\frac{\lceil \log (\hat{\varepsilon}/\check{\varepsilon}) \rceil}{\beta}\right)$
    
    \If{$\widetilde{Q}_{\text{\emph{count}}}(D \cap \mathcal{X}_i) \geq T_i$}{
        $\ell \gets i$
        
        \textbf{break}
    }
}

$\varepsilon_\tau \gets 2^{\ell-1} \cdot \check{\varepsilon}$

\tcp*[l]{Aggregate results from the remaining domains}
$\widetilde{Q}_{\text{count}}(D) \gets \sum\limits_{i=\ell}^{\lceil \log \frac{\hat{\varepsilon}}{\check{\varepsilon}} \rceil} \widetilde{Q}_{\text{count}}(D \cap \mathcal{X}_i)$

\Return $\varepsilon_\tau, \, \widetilde{Q}_{\text{count}}(D)$
\end{algorithm}

Now, we analyze the privacy and utility of Algorithm~\ref{alg:cnt}.  
For privacy, on any given domain $\mathcal{X}_i$, the algorithm is executed with the minimum privacy budget within the entire domain, ensuring $\frac{1}{2^{i-1} \cdot \check{\varepsilon}}$-DP.  
Therefore, the mechanism also satisfies $\mathcal{E}$-PrDP on $\mathcal{X}_i$.  
Notably, any individual record can only affect a single $\mathcal{X}_i$ due to the disjoint partitioning.  
By the properties of parallel composition (Lemma~\ref{lem:para}), the entire mechanism satisfies $\mathcal{E}$-PrDP.

For utility, Theorem~\ref{the:cnt} provides the error bound for both the counting results and $\ell$. 
We present the proof of the utility of Algorithm~\ref{alg:cnt}, as stated in Theorem~\ref{the:cnt}.

\begin{theorem}
\label{the:cnt}
    For any given privacy budget function $\mathcal{E}(\cdot)$ and $\beta \in (0,1)$, Algorithm~\ref{alg:cnt} returns $\varepsilon_\tau$ and $\widetilde{Q}_{\text{count}}(D)$, such that with probability at least $1 - \beta$, we have $\varepsilon_\tau \geq \frac{1}{2} \varepsilon_{\min}(D)$,
    and
    \[\big|\widetilde{Q}_{\text{\emph{count}}}(D)-Q_{\text{\emph{count}}}(D)\big| = O\left(\frac{1}{\varepsilon_{\min}(D)} \log\left(\frac{ \log (\hat{\varepsilon} / \check{\varepsilon}) }{\beta}\right)\right).\]
\end{theorem}

\begin{proof}
\label{prf:cnt_pri_bnd}  
We analyze the error introduced by the algorithm, which consists of two components.  

The first component arises from omitting the results of the first $\ell-1$ privacy-specified domains.  
According to Lemma~\ref{lem:tail} and the union bound over $\lceil \log (\hat{\varepsilon} / \check{\varepsilon}) \rceil$ domains, with probability at least $1 - \beta$, we have  
\begin{equation}
\label{eqn:err_bd}
    \begin{aligned}
    |\tilde{Q}_{\text{count}}(D \cap \mathcal{X}_i) - Q_{\text{count}}(D \cap \mathcal{X}_i)| &\leq T_i \\
     &= \frac{1}{2^{i-1} \cdot \check{\varepsilon}} \ln\left(\frac{\lceil \log (\hat{\varepsilon}/\check{\varepsilon}) \rceil}{\beta}\right),
    \end{aligned}
\end{equation}
which holds for all domains. 
This means Algorithm \ref{alg:cnt} will not stop at empty domains with $Q_{\text{count}}(D \cap \mathcal{X}_i) = 0$, to be more specific, we have
\begin{equation}
\label{eqn:bound_eps}
    \varepsilon_\tau \geq \frac{1}{2} \varepsilon_{\text{min}}(D).
\end{equation}
(\ref{eqn:err_bd}) also indicates that for each domain with $i < \ell$, the number of elements within those domains is very limited even if they are discarded  
\begin{equation*}
    Q_{\text{count}}(D \cap \mathcal{X}_i) \leq 2 \cdot \frac{1}{2^{i-1} \cdot \check{\varepsilon}} 
    \ln\left(\frac{\lceil \log (\hat{\varepsilon} / \check{\varepsilon}) \rceil}{\beta}\right),
\end{equation*} 
That is, for domains preceding $\ell$, they are either empty—if they are before the domain containing $\varepsilon_{\min}(D)$—or very ``light'' if they coincide with or follow the domain containing $\varepsilon_{\min}(D)$:
\begin{equation}
\begin{aligned}
\label{eqn:part1}
    \sum_{i=1}^{\ell-1} Q_{\text{count}}(D \cap \mathcal{X}_i) 
    &= O\left(\frac{1}{\varepsilon_{\min}(D)} 
    \log\left(\frac{\lceil \log (\hat{\varepsilon} / \check{\varepsilon}) \rceil}{\beta}\right)\right) \\
    &= \tilde{O}\left(\frac{1}{\varepsilon_{\min}(D)}\right).
\end{aligned}
\end{equation}
This part achieves sub-goal (a).  

The second component comes from the intrinsic error of the Laplace mechanism.  
According to (\ref{eqn:bound_eps}), the cumulative error over the retained domains is bounded by  
\begin{equation}
\label{eqn:part2}
    \begin{aligned}
        &\sum_{i=\ell}^{\lceil \log (\hat{\varepsilon} / \check{\varepsilon}) \rceil} 
        |\tilde{Q}_{\text{count}}(D \cap \mathcal{X}_i) - Q_{\text{count}}(D \cap \mathcal{X}_i)| \\
        \leq & \sum_{i=\ell}^{\lceil \log (\hat{\varepsilon} / \check{\varepsilon}) \rceil} \frac{1}{2^{i-1} \cdot \check{\varepsilon}} \ln\left(\frac{\lceil \log (\hat{\varepsilon} / \check{\varepsilon}) \rceil}{\beta}\right) \\
        \leq & O\left(\frac{1}{\varepsilon_\tau} 
        \log\left(\frac{\lceil \log (\hat{\varepsilon} / \check{\varepsilon}) \rceil}{\beta}\right)\right) \\
        \leq & O\left(\frac{1}{\varepsilon_{\text{min}}(D)} 
        \log\left(\frac{\lceil \log (\hat{\varepsilon} / \check{\varepsilon}) \rceil}{\beta}\right)\right) \\
        = & \tilde{O}\left(\frac{1}{\varepsilon_{\text{min}}(D)}\right).
    \end{aligned}
\end{equation}
From the second to the third line, we observe that the series follows an exponential pattern.  
Therefore, the summation is on the order of its first term.  
This bound satisfies sub-goal (b) discussed earlier.  

By combining (\ref{eqn:part1}) and (\ref{eqn:part2}), we obtain the desired error bound stated in Theorem~\ref{the:cnt}.
\end{proof}



\begin{example}
Suppose we have a bank dataset with a million records and aim to count customers with a specific mailing postcode. Given the privacy budget function in (\ref{eqn:pbf_exp}) with $\alpha = 10^4$, $\check{\varepsilon} = 7.8125\times10^{-6}$, and $\hat{\varepsilon} = 4.096$, i.e., we set $U = 1,280,000,000$~\$.
Figure~\ref{fig:alg_cnt} illustrates how Algorithm~\ref{alg:cnt} works. 
It first computes noisy counts over all domains and compares them with their respective thresholds.  
Then, it identifies $\mathcal{X}_{12}$ as the first sub-domain with $\widetilde{Q}_{\text{count}}(D \cap \mathcal{X}_{12}) = 1729.38 > T_{12} = 437.69$. Therefore, the final result is
\begin{equation*}
\begin{aligned}
\widetilde{Q}_{\text{count}}(D) &= \sum_{i=12}^{19} \widetilde{Q}_{\text{count}}(D \cap \mathcal{X}_i) \\
&= 1729.38 + 4042.73 + 2384.80 \\
&\quad + 1042.90 + 484.86 + 199.22 + 92.26 - 0.14 \\
&= 9976.01.
\end{aligned}
\end{equation*}
The ground truth result for this dataset is $\sum_{i=1}^{19} Q_{\text{count}}(D \cap \mathcal{X}_i) = 10,000$.   
Additionally, we return an $\varepsilon_{\tau} = 0.016$. \qed

\end{example}


\subsection{Extensions beyond Basic Counting}
In this section, we extend Algorithm~\ref{alg:cnt} to two classic problems: sum estimation and distinct count.  
These extensions offer deeper insights into the algorithm's strengths and limitations, inspiring a more general framework that eliminates the need for query-specific designs and supports diverse queries and privacy budget functions.

\paragraph{Sum Estimation}
\label{sec:ext-sum}

For sum estimation, the SOTA mechanism under standard $\varepsilon$-DP achieves an error of $\tilde{O}(\text{Max}(D)/\varepsilon)$.  
Our goal is to achieve $\tilde{O}(\text{Max}(D)/\varepsilon_{\min}(D))$ in the PrDP setting.
To extend Algorithm~\ref{alg:cnt} for sum estimation, we perform a privacy-specified domain partition and estimate the sum within each sub-domain by adding Laplace noise with a scale proportional to 
$\max_{r \in \mathcal{X}_i} \left( \frac{v_{\text{Bal}}}{\mathcal{E}(r)} \right)$.  
We then identify the first sub-domain where 
$\widetilde{Q}_{\text{sum}}(D \cap \mathcal{X}_i) \geq \max_{r \in \mathcal{X}_i} \left( \frac{v_{\text{Bal}}}{\mathcal{E}(r)} \right) \ln\left(\frac{\lceil \log (\hat{\varepsilon} / \check{\varepsilon}) \rceil}{\beta}\right)$.  
Finally, we return the sum over the domains from the $\ell$-th to the $\lceil \log (\hat{\varepsilon} / \check{\varepsilon}) \rceil$-th.

For specific cases, this approach can achieve the desired error bound.  
For bank deposit estimation with $\mathcal{E}(r) = \alpha / v_{\text{Bal}}$, the error bound is $\tilde{O}(\text{Max}(D)/\varepsilon_{\min}(D))$ with constant probability. 
However, this approach does not guarantee good utility for all privacy budget functions.  
For example, when we use $\mathcal{E}(r) = \alpha/ \log (v_{\text{Bal}})$, the error bound changes to $\tilde{O}(\text{Max}(D)^2/\varepsilon_{\min}(D))$. 
This is because, when we partition the domains according to privacy budget, each privacy-specified sub-domain $\mathcal{X}_\ell$ now covers a much wider value range.
For example, when we consider the sub-domain that $\text{Max}(D)$ is located in, the maximum balance in that sub-domain can be as large as $\text{Max}(D)^2$. 
Then, the noise added to the sum result of that sub-domain must be at the scale of $\tilde{O}(\text{Max}(D)^2 / \varepsilon_{\min}(D))$.
For a detailed discussion and proof, see Appendix~A of the full version~\cite{full_version}.

\paragraph{Non-Union Preserving Queries}

More generally, the idea of Algorithm~\ref{alg:cnt} cannot handle non-union-preserving queries, namely, those for which $Q\left(\bigcup_i D_i\right) \neq \sum_i Q(D_i)$, which is a common scenario in practice.  
Examples include distinct count, median, maximum, mode estimation, or join counting queries in relational databases. 
Such queries cannot be decomposed into sub-queries over privacy-specified domains $\mathcal{X}_i$ and summed.  
To be more precise, consider a bank determining the number of districts its customers reside in, identified by $v_{\text{PC}}$. This reduces to counting distinct $v_{\text{PC}}$ in $D$.  
If the privacy budget depends on multiple factors, i.e., $\mathcal{E}(r) = \mathcal{E}(v_{\text{Bal}}, \, v_{\text{PC}}, \, \ldots)$, the same postcode may appear in multiple privacy-specified domains, leading to duplicate counts and non-additive results, making partitioning inapplicable.

Despite these challenges, the following sections will show how insights from the basic counting algorithm remain valuable. 
These insights form the foundation of a general framework that supports all query types under PrDP without constraining the privacy budget function's form.
    \section{A General Framework for PrDP}
\label{sec:gen_fmk_PrDP}

\begin{figure*}[t]
    \centering
    \includegraphics[width=0.9\textwidth]{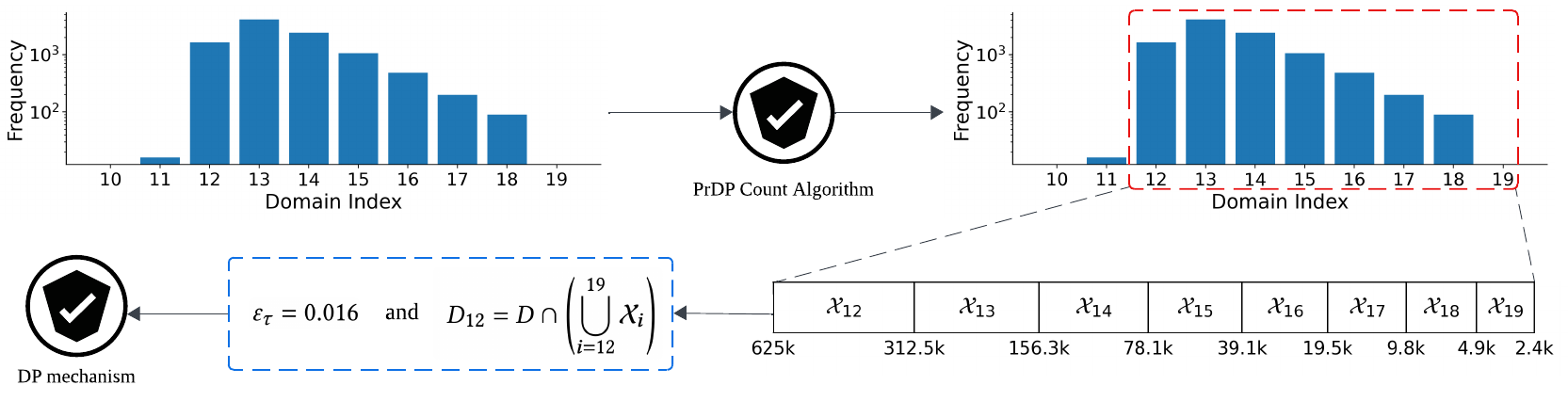}
    \caption{Illustration of the workflow of Algorithm~\ref{alg:gfra}.}
    \label{fig:alg_gen}
\end{figure*}

We have shown that the basic counting problem under PrDP can be effectively solved through privacy-specified domain partitioning, achieving the desired error bound of $\tilde{O}(1/\varepsilon_{\text{min}(D)})$. 
In this section, we propose a general framework that handles both union and non-union-preserving queries while accommodating any given privacy budget function.
\revone{Roughly speaking, the general PrDP framework first uses our Laplace-based PrDP counting mechanism to accurately estimate $\varepsilon_{\min}(D)$.
It then employs any existing standard DP mechanism with the estimated privacy budget to answer the desired query.}
Moreover, our framework achieves an instance-specific error bound by reducing the error dependency on the global parameter $\check{\varepsilon}$ to $\varepsilon_{\min}(D)$ for $D$.


We follow the intuition of Algorithm~\ref{alg:cnt} but split the process into two steps, each consuming half of the privacy budget.  
In the first step, we estimate a reasonable threshold $\varepsilon_\tau$ that is not too small while ensuring that most records' privacy budgets fall within the interval $[\varepsilon_\tau, \, \hat{\varepsilon}]$.  
This is achieved by invoking Algorithm~\ref{alg:cnt} with the privacy function  
\begin{equation}
\label{eqn:half_pbf}
    \mathcal{E}'(r) := \frac{1}{2} \cdot \mathcal{E}(r).
\end{equation}
After this step, we obtain $\varepsilon_\tau = 2^{\ell-1} \cdot \check{\varepsilon}$ under $\mathcal{E}$'-PrDP, satisfying the following properties:
\begin{enumerate}[label=(\roman*)]
    \item $\left|D \cap \left(\bigcup_{i=1}^{\ell-1} \mathcal{X}i\right) \right| \leq \tilde{O}(1/\varepsilon_{\min}(D))$
    \item $\varepsilon_\tau \geq \frac{1}{2} \varepsilon_{\min}(D)$
\end{enumerate}

Then, instead of answering the query on each privacy-specified domain separately and summing up the results, in the second step, we treat the remaining data as a whole.  
Specifically, we union all sub-domains with domain index larger than $\ell$, then invoke  
a $\frac{\varepsilon_\tau}{2}$- (standard) DP mechanism $\mathcal{M}$ over  
$\bigcup_{i=\ell}^{\lceil \log (\hat{\varepsilon} / \check{\varepsilon}) \rceil} \mathcal{X}_i$.  
For simplicity, we denote $D \cap \left(\bigcup_{i=\ell}^{\lceil \log(\hat{\varepsilon}/\check{\varepsilon}) \rceil} \mathcal{X}_i\right)$ as $D_\ell$.  
\revone{Excluding those ``light'' domains with overly strong privacy protection needs is critical to our general framework, as it balances utility and privacy to avoid error dominated by $\check{\varepsilon}$, and instead depends only on $\varepsilon_{\min}(D)$.}
Since the remaining privacy budget is at least $\frac{\varepsilon_\tau}{2}$ across the entire unioned privacy-specified domain, the second step also preserves $\mathcal{E}'$-PrDP.  

Please note that the choice of $\mathcal{M}$ is at the discretion of the data curator.  
They can choose any suitable existing DP mechanism or design a custom one to meet diverse real-world requirements.

\revtwo{This applies to all types of SJA queries, meaning that any standard DP mechanism for such queries can benefit from our framework. 
However, due to the complex error characteristics of these mechanisms~\cite{dong21:residual, dong2021nearly,johnson2018towards, tao2020computing,dong2023better,cai2023privlava} (see a survey paper for a comprehensive study~\cite{dong2023query}), it is not straightforward to derive formal error guarantees for the resulting PrDP solution. 
In conclusion, our framework serves as a transparent, glass-box transformation that adapts standard DP mechanisms to PrDP settings.} 
\revone{It fails to answer a query only when no standard DP mechanism exists for that query.}

The detailed algorithm is presented in Algorithm~\ref{alg:gfra}.  
Additionally, Figure~\ref{fig:alg_gen} illustrates the workflow of the proposed general PrDP framework.
The runtime complexity of the framework is primarily determined by the complexity of the selected standard DP mechanism, since estimating $\varepsilon_\tau$ consumes only $O(n)$.

For privacy, by the sequential composition rule in Lemma~\ref{lem:seq}, the entire framework satisfies  
$2 \cdot \mathcal{E}'$-PrDP, which simplifies to $\mathcal{E}$-PrDP.
For the utility, property (i) means at most $\tilde{O}(1/\varepsilon_{\text{min}}(D))$ records will be excluded.
Property (ii) implies when we call the $\frac{\varepsilon_\tau}{2}$- (standard) DP mechanism $\mathcal{M}$ in the second step, the error will \revtwo{scale} with $\varepsilon_{\text{min}}(D)$.

\begin{algorithm}[t]
\LinesNumbered
\caption{General PrDP Framework}
\label{alg:gfra}
\KwIn{Privacy budget function $\mathcal{E}(\cdot)$, dataset $D$, failure rate $\beta$, DP mechanism $\mathcal{M}$}
\KwOut{$\tilde{Q}(D)$ under $\mathcal{E}$-PrDP}

\tcp*[l]{Part 1: Estimate $\varepsilon_\tau$}

$\varepsilon_\tau \gets \text{Algorithm}~\ref{alg:cnt},\,\text{PrDP Count}\left(\,\mathcal{E}'(r),\, D,\, \tfrac{\beta}{2}\right)$

\tcp*[l]{Part 2: Execute the DP mechanism}

$\tilde{Q}(D) \gets \mathcal{M}\left(
    D \cap \left(\bigcup_{i=\ell}^{\lceil \log(\hat{\varepsilon}/\check{\varepsilon}) \rceil} \mathcal{X}_i\right),\,
    \frac{1}{2}\,\varepsilon_\tau,\,
    \tfrac{\beta}{2}
\right)$

\Return $\tilde{Q}(D)$
\end{algorithm}

\begin{theorem}
\label{the:fmk_bound}
    For any given privacy budget function $\mathcal{E}(\cdot)$ and $\beta \in (0,1)$, as well as any DP mechanism $\mathcal{M}$ for query $Q$,
    Algorithm~\ref{alg:gfra} returns $\widetilde{Q}(D)$, such that with probability at least $1 - \beta$, we have
       
    \begin{equation*}
        |\widetilde{Q}(D) - Q(D) | \leq \underbrace{|Q(D) - Q(D')|}_{\text{I}} + \underbrace{\text{\emph{Err}}_{\mathcal{M}}\left(D',\, \varepsilon_{\min}(D)/4,\, \beta/2\right)}_{\text{II}},
    \end{equation*}
    where $D'$ is obtained by removing  
    $O\left(\frac{1}{\varepsilon_{\min}(D)} \log \frac{ \log (\hat{\varepsilon} / \check{\varepsilon}) }{\beta} \right)$  
    records from $D$ and $\text{Err}_{\mathcal{M}}(\cdot,\, \cdot,\, \cdot)$ denotes the error introduced by the DP mechanism  
    $\mathcal{M}$, which depends on the input dataset, privacy budget, and failure rate.    
\end{theorem}

\begin{proof}
    Firstly, according to Theorem~\ref{the:cnt}, we know that  
    \begin{align*}
        \label{eqn:trim_bnd}
        \big| D \setminus D_\ell \big| &\leq O\left(\frac{1}{\frac{1}{2} \varepsilon_{\min}(D)} \log \frac{2 \lceil \log (\hat{\varepsilon} / \check{\varepsilon}) \rceil}{\beta} \right)  \notag \\
        &= O\left(\frac{1}{\varepsilon_{\min}(D)} \log \frac{\lceil \log (\hat{\varepsilon} / \check{\varepsilon}) \rceil}{\beta} \right).
    \end{align*}   
    Thus, $D_\ell$ itself satisfies the requirement for $D'$.
    
    As for the error of the selected mechanism $\mathcal{M}$ over the remaining records $D'$,  
    its magnitude is primarily determined by $\varepsilon_{\min}(D)$.  
    According to Theorem~\ref{the:cnt}, we have $\varepsilon_\tau \geq \frac{1}{2} \varepsilon_{\min}(D)$.  
    Therefore, part II holds.

    Combining these two parts, we derive the stated error bound in Theorem~\ref{the:fmk_bound}.  

    Finally, the probability that the bound holds with at least $1 - \beta$ follows from the union bound over the two consecutive algorithms, each of which guarantees utility with probability at least $1 - \frac{\beta}{2}$.  
\end{proof}

As shown in the above theorem, the error consists of two components:  
\begin{enumerate}[label=\Roman*]
    \item comes from the bias of excluding $\tilde{O}(1/\varepsilon_{\min}(D))$ records.  
    \item comes from the error of invoking the DP mechanism on the remaining records with $\varepsilon_\tau/2$.
\end{enumerate}
We now provide intuition showing that, in most cases, these two components are not significantly larger than
\begin{equation}
\label{eqn:fmk_target}
    \text{Err}_{\mathcal{M}}\left(D,\, \varepsilon_{\min}(D)/4,\, \beta/2\right),
\end{equation}
which is our target. 
While this is not a formal guarantee, we will further demonstrate it holds in most common queries such as sum estimation and distinct count.

First, following the notation in Theorem~\ref{the:dn_bound}, component I is exactly $\Delta Q^{\rho}(D)$ with $\rho = \tilde{O}(1/\varepsilon_{\min}(D))$.  
As discussed in Section~\ref{sec:opt_DP}, under standard DP, most existing mechanisms have an error at least $\Omega(\Delta Q^{\rho}(D))$ with $\rho = \tilde{O}(1/\varepsilon_{\min}(D))$ if we use a privacy budget $\varepsilon=\varepsilon_{\min}(D)$. 
Therefore, component I will not be significantly larger than (\ref{eqn:fmk_target}), at the very least.

Moving towards component II, first, many standard DP mechanisms achieve an error that is independent of the input dataset. 
Additionally, for common queries such as sum estimation and join counting, existing standard DP mechanisms achieve an instance-specific error that is supermodular with respect to the input dataset $D$.  
Intuitively, adding more records does not reduce query sensitivity and, therefore, does not facilitate achieving DP, implying that component II should not exceed (\ref{eqn:fmk_target}). 
Moreover, in other cases where the error function is instance-specific but not supermodular, the difference between datasets $D$ and $D'$ remains limited, ensuring that the errors incurred on $D$ and $D'$ are still very similar.




\paragraph{Sum estimation}
As mentioned in Section~\ref{sec:ext-sum}, the naive extension fails to attain the desired error bound $\tilde{O}(\text{Max}(D)/\varepsilon_{\min}(D))$ for certain privacy budget functions like $\mathcal{E}(r) = \alpha'/v_{\text{bal}}$.  
Instead, for Algorithm~\ref{alg:gfra}, component I is $\tilde{O}\left(\text{Max}(D)/\varepsilon_{\min}(D)\right)$.  
For component II, invoking \cite{dong2023universal} with a privacy budget over $D'$ results in an error of $\tilde{O}(\text{Max}(D')/\varepsilon_{\min}(D))$, which is further bounded by $\tilde{O}(\text{Max}(D)/\varepsilon_{\min}(D))$.  
Overall, the final error is $\tilde{O}\left(\text{Max}(D)/\varepsilon_{\min}(D)\right)$, aligning with our target.


\paragraph{Distinct count}  
For distinct count, the standard DP mechanism has an error of $O(1/\varepsilon_{\min}(D))$, therefore, our target under PrDP is $\tilde{O}(1/\varepsilon_{\min}(D))$.  
For Algorithm~\ref{alg:gfra}, component I is $\tilde{O}(1/\varepsilon_{\min}(D))$.  
Meanwhile, component II is $O(1/\varepsilon_{\min}(D))$.  
Combining them together gives us the desired $\tilde{O}(1/\varepsilon_{\min}(D))$.



    \section{Per-Record Local Differential Privacy}
\label{sec:PrLDP}

In this section, we extend our framework to the \textit{local differential privacy} (LDP) model \cite{mcgregor2010limits, bassily2015local}, where we achieve an error with a dependency on $\varepsilon_{\min}(D)$, similar to that in Section~\ref{sec:gen_fmk_PrDP}.

\subsection{Per-record Local Differential Privacy}

Let us define per-record LDP (PrLDP) first. To begin with, we revisit the definition of LDP.
Recall that under central DP, there is a trusted data curator to collect and process data.
Under LDP, there is no trusted curator. 
Instead, each party processes their record locally and then sends the result to the analyzer directly.
During this process, each party invokes an LDP protocol to protect the privacy of their record.
The formal definition is given below:

\begin{definition}
\label{def:LDP}
    \emph{($\varepsilon$-Local Differential Privacy)}.  
    For a given $\varepsilon > 0$,  $\mathcal{M}$ is $\varepsilon$-local differential privacy ($\varepsilon$-LDP) if for any pair $r,\, r' \in [U]^{d}$ and for all possible outputs $y$, the following holds:
    \begin{equation*}
        \Pr[\mathcal{M}(r) = y] \leq e^\varepsilon \cdot \Pr[\mathcal{M}(r') = y].
    \end{equation*}
\end{definition}

This definition ensures that the outputs of the mechanism $\mathcal{M}$ for any pair of records are $\varepsilon$-indistinguishable, thereby providing strong privacy protection for each individual record.
\revone{Consider the bank deposit estimation scenario in which customers no longer trust the bank to process their records. 
Each customer therefore performs local processing to privatize their data before sending them to the bank. 
Because privacy requirements may vary across customers, it is natural to extend Definition~\ref{def:LDP} to a per-record privacy budget, whereby the privacy budget for each customer is determined by a shared privacy budget function—e.g., the function shown in~(\ref{eqn:pbf_exp}).} 
This introduces an additional challenge: determining whether the privacy function should take $r$ or $r'$.
Such an issue does not exist in our PrDP model, as the central DP model defines neighboring datasets using the add-delete-one paradigm, where one dataset is obtained by inserting or deleting a single record from the other.
In contrast, the LDP model uses the change-one paradigm, where the record can be changed to an arbitrary one.
Under standard DP, it is well established that these two notions provide equivalent levels of privacy protection up to small constant factors. 
Specifically, any change-one pair of datasets can be seen as differing by a distance of two under the add-delete-one definition: deleting one record and inserting another. 
Conversely, by introducing a dummy record $\perp$ that has no influence on the computation, one can simulate the insertion or deletion of a record as a change: replacing a real record with $\perp$, or vice versa.

Following this idea, we define PrLDP under the add-delete-one paradigm using the dummy record $\perp$:
\begin{definition}
\label{def:prldp}
    \emph{(Per-record Local Differential Privacy)}.  
    Given a record-dependent privacy budget function $\mathcal{E}: [U]^{d} \to [\check{\varepsilon}, \, \hat{\varepsilon}]$, a mechanism $\mathcal{M}$ is $\mathcal{E}$-per-record local DP, or simply $\mathcal{E}$-PrLDP, if for arbitrary input record $r \in [U]^{d}$ and for all possible outputs $y$, the following holds:
    \begin{equation*}
        e^{-\mathcal{E}(r)} \Pr[\mathcal{M}(\perp) \in y] \leq \Pr[\mathcal{M}(r) \in y] \leq e^{\mathcal{E}(r)} \Pr[\mathcal{M}(\perp) \in y].
    \end{equation*}
\end{definition}
For specific queries, we may replace $\perp$ with a meaningful default value. 
For example, in counting and summation queries, we can set $Q(\perp) := 0$.

\paragraph{Unique challenge of PrLDP}
Compared with PrDP, the main challenge of PrLDP is that we need to add noise to each data record.
However, since the privacy budget used depends on the value of the record, the scale of the added noise can inadvertently leak information about the record itself.
This also highlights the challenge of PrDP over PDP.
As previously mentioned, a fundamental distinction between PrDP and PDP lies in that, under PDP, the privacy budgets assigned to individual records are public, making it safe to release record-specific noise. 
In contrast, PrDP treats these budgets as private.

\subsection{Privacy-Specified Query Augmentation for Counting under PrLDP}
\label{sec:prldp_cnt}
We first consider the counting problem. 
Existing LDP protocol achieves an error of $O(\sqrt{n}/\varepsilon)$, where $n$ is the number of parties and each party adds noise with scale $O(1/\varepsilon)$ to their count.
Moving to PrLDP, as previously mentioned, utilizing record-specified noise will lead to privacy leakage.
To address this issue, we propose a \textit{privacy-specified query augmentation} method.
More specifically, we partition the interval $[\check{\varepsilon},\, \hat{\varepsilon}]$ into $\lceil \log(\hat{\varepsilon}/\check{\varepsilon}) \rceil$ sub-intervals, which correspond to $\lceil \log(\hat{\varepsilon}/\check{\varepsilon}) \rceil$ privacy-specified domains. 
For the $i$-th domain, we define the query on a data record $r$ as
\begin{equation*}
    Q_{\text{count},\, i}(r) = \mathbb{I}(r\in \mathcal{X}_i),
\end{equation*}
where $\mathbb{I}(\cdot)$ denotes the indicator function.
Note that $\perp$ does not belong to any domain. 
Thus, $Q_{\text{count},\, i}(\perp) \equiv 0$.

\revtwo{Then, the local randomizer further privatizes the $\lceil \log(\hat{\varepsilon}/\check{\varepsilon}) \rceil$ real counts by adding Laplace noise with scale $1/(2^{i-1} \cdot \check{\varepsilon})$}
\begin{equation*}
    \widetilde{Q}_{\text{count},\, i}(r) = Q_{\text{count},\, i}(r) + \text{Lap}\left(\frac{1}{2^{i-1} \cdot \check{\varepsilon}}\right).
\end{equation*}

Overall, each party sends $\lceil \log(\hat{\varepsilon}/\check{\varepsilon}) \rceil$ noisy counting results to augment their response.
After receiving responses from all $n$ parties, the data analyzer aggregates the results for each privacy-specified domain and compares them against the threshold
\begin{equation*}
    T_i := \sqrt{8n} \cdot \frac{1}{2^{i-1} \cdot \check{\varepsilon}} \ln\left(\frac{\lceil \log(\hat{\varepsilon} / \check{\varepsilon}) \rceil}{\beta}\right).
\end{equation*}
Finally, similar to the process in Algorithm~\ref{alg:cnt}, it identifies the first domain, $\mathcal{X}_l$, whose result exceeds the corresponding threshold, and sums the responses over all domains with $i \geq l$ as the final result.
The detailed algorithms for the local randomizer and the data analyzer are presented in Algorithm~\ref{alg:local_prldp_count} and Algorithm~\ref{alg:analyzer_prldp_count}, respectively.

For privacy, each data record 
affects only a single domain, which is protected by noise calibrated to the strongest privacy requirement of that domain. 
By using parallel composition across the privacy-specified domains, our privacy-specified query augmentation approach satisfies PrLDP.
Regarding utility, Theorem~\ref{the:ldpcnt} provides theoretical guarantees for both $\widetilde{Q}_{\text{count}}(D)$ and $\varepsilon_\tau$.

\begin{algorithm}[t]
\LinesNumbered
\caption{PrLDP Count (Randomizer)}
\label{alg:local_prldp_count}
\KwIn{Privacy budget function $\mathcal{E}(r)$, record $r_j$}
\KwOut{$\{\widetilde{Q}_{\text{count},\, i}(r_j)\}_{i=1}^{\lceil \log(\hat{\varepsilon}/\check{\varepsilon})\rceil}$ under PrLDP}

\For{$i \gets \{1, \, 2, \, \ldots, \, \lceil \log(\hat{\varepsilon}/\check{\varepsilon}) \rceil\}$}{
    $\widetilde{Q}_{\text{count},\, i} (r_j) \gets Q_{\text{count}, i} (r_j)+ \text{Lap}\left(\frac{1}{2^{i-1} \cdot \check{\varepsilon}}\right)$
}

\textbf{Send:} $\{\widetilde{Q}_{\text{count},\, i}(r_j)\}_{i=1}^{\lceil \log(\hat{\varepsilon}/\check{\varepsilon})\rceil}$
\end{algorithm}

\begin{algorithm}[h]
\LinesNumbered
\caption{PrLDP Count (Data Analyzer)}
\label{alg:analyzer_prldp_count}
\KwIn{Noisy answers $\{\{\widetilde{Q}_{\text{count},\, i}(r_j)\}\}_{i=1}^{\lceil \log(\hat{\varepsilon}/\check{\varepsilon})\rceil}\}_{j=1}^n$, failure rate $\beta$}
\KwOut{$\widetilde{Q}_{\text{count}}(D)$ under PrLDP}

$\ell \gets \left\lceil \log\left(\frac{\hat{\varepsilon}}{\check{\varepsilon}}\right)\right\rceil$

\tcp*[l]{Obtain results}

\For{$i \gets \{1,2,\ldots,\lceil \log(\hat{\varepsilon}/\check{\varepsilon}) \rceil\}$}{

    $\widetilde{Q}_{\text{count}, \, i} \gets \sum_{j=1}^{n}
    \widetilde{Q}_{\text{count},\, i}(r_j)$
}

\tcp*[l]{Estimate $\ell$, i.e., $\varepsilon_\tau$}

\For{$i \gets \{1,2,\ldots,\lceil \log(\hat{\varepsilon}/\check{\varepsilon}) \rceil\}$}{
    $T_i \gets \sqrt{8n}\,\cdot\,\frac{1}{2^{i-1}\,\check{\varepsilon}} \,\ln\left(\frac{\lceil \log(\hat{\varepsilon}/\check{\varepsilon}) \rceil}{\beta}\right)$

    \If{$\widetilde{Q}_{\text{\emph{count}}, \, i} \geq T_i$}{
        $\ell \gets i$

        \textbf{Break}
    }
}

$\varepsilon_\tau \gets 2^{l-1} \cdot \check{\varepsilon}$


\tcp*[l]{Aggregate results from the remaining domains}

$\widetilde{Q}_{\text{count}}(D) \gets \sum_{i=\ell}^{\lceil \log\left(\frac{\hat{\varepsilon}}{\check{\varepsilon}}\right)\rceil} \widetilde{Q}_{\text{count}, \, i}$

\Return $\varepsilon_\tau, \, \widetilde{Q}_{\text{count}}(D)$
\end{algorithm}

\begin{theorem}
\label{the:ldpcnt}
    For any given privacy budget function $\mathcal{E}(\cdot)$ and $\beta \in (0,1)$, the combination of Algorithm~\ref{alg:local_prldp_count} and Algorithm ~\ref{alg:analyzer_prldp_count} returns $\varepsilon_\tau$ and $\widetilde{Q}_{\text{count}}(D)$, such that with probability at least $1 - \beta$, we have $\varepsilon_\tau \geq \frac{1}{2} \varepsilon_{\min}(D)$,
    and
    \begin{equation*}
        \big|\widetilde{Q}_{\text{\emph{count}}}(D)-Q_{\text{\emph{count}}}(D)\big| = O\left(\frac{\sqrt{n}}{\varepsilon_{\min}(D)} \log\left(\frac{ \log (\hat{\varepsilon} / \check{\varepsilon}) }{\beta}\right)\right).
    \end{equation*}
\end{theorem}

\begin{proof}
    Algorithm~\ref{alg:analyzer_prldp_count} satisfies PrLDP by Lemma~\ref{lem:post}.  
    For utility, the analysis follows similarly to the proof of Theorem~\ref{the:cnt}, where the $\sqrt{n}$ factor arises from aggregating $n$ noisy responses—standard in LDP settings.
    More precisely, for $n$ independent and identically distributed Laplace noises, the variance of their sum increases by a factor of $n$, resulting in a larger error magnitude proportional to $\sqrt{n}$.
\end{proof}

\subsection{A General Framework for PrLDP}
\label{sec:prldp_fmk}
In the previous section, we introduced a counting algorithm under PrLDP based on the privacy-specified query augmentation technique.  
Following a similar strategy as the centralized setting, it is not hard to build a general framework for extending existing LDP protocol to the PrLDP model.

Our protocol involves two rounds.
Specifically, in the first round, we use Algorithm~\ref{alg:local_prldp_count} and Algorithm~\ref{alg:analyzer_prldp_count} to estimate $\varepsilon_\tau$ with privacy budget function $\mathcal{E}'(r) := \frac{1}{2} \mathcal{E}(r)$.
In the second round, we exclude records whose privacy budgets fall below $\varepsilon_\tau$.  
More precisely, each party will replace its data record with $\perp$ if that record corresponding to a privacy budget less than $\varepsilon_{\tau}$. 
Then, each party invokes a LDP protocol with a privacy budget of $\frac{\varepsilon_\tau}{2}$. 
The details are provided in Algorithm~\ref{alg:prldp_general}.
\revtwo{Roughly speaking, similar to the general framework for PrDP, the PrLDP general framework first uses our PrLDP counting protocol to obtain $\varepsilon_\tau$, and then sets the records with privacy budgets below this threshold to dummy records. 
Finally, it applies an existing LDP protocol to answer the desired queries. 
Therefore, it supports any query for which an LDP protocol exists, regardless of whether it is union-preserving or non-union-preserving. 
For example, since no existing LDP protocol supports distinct count, our framework does not support this type of query.}

For privacy, by Lemma~\ref{lem:para}, the first round of computation preserves $\frac{1}{2}\mathcal{E}$-PrLDP.
In the second round, if a party's record corresponds to a privacy budget less than $\varepsilon_\tau$, setting it to $\perp$ prevents any privacy leakage. 
Otherwise, the privacy loss is bounded by $\frac{\varepsilon_\tau}{2}$. Therefore, the second round also satisfies $\frac{1}{2}\mathcal{E}$-PrLDP.
By the composition theorem, the overall framework satisfies $\mathcal{E}$-PrLDP.

\begin{algorithm}[htbp]
\LinesNumbered
\caption{General PrLDP Framework}
\label{alg:prldp_general}
\KwIn{Privacy budget function $\mathcal{E}(\cdot)$, records $\{r_j\}_{j=1}^{n}$, failure rate $\beta$, LDP mechanism $\mathcal{M_{\text{local}}}$, $\mathcal{M_{\text{curator}}}$}
\KwOut{$\widetilde{Q}(D)$ under PrLDP}

\tcp*[l]{Round I: estimate $\varepsilon_\tau$}

\For{$j \gets \{1,2,\ldots,n\}$}{
    $\{\widetilde{Q}_{\text{count},\, i}(r_j)\}_{i=1}^{\lceil \log(\hat{\varepsilon}/\check{\varepsilon})\rceil} \gets \text{randomizer}\left(\mathcal{E}'(r),\, r_j\right)$
}  

$\varepsilon_\tau \gets \text{analyzer}\left(\{\{\widetilde{Q}_{\text{count},\, i}(r_j)\}\}_{i=1}^{\lceil \log(\hat{\varepsilon}/\check{\varepsilon})\rceil}\}_{j=1}^n, \, \tfrac{\beta}{2} \right)$

\tcp*[l]{Round II: execute the LDP protocal}

\For{$j \gets 1$ \textbf{to} $n$}{

\tcp*[l]{$r$ with small privacy budget respond as $\perp$}

    $\widetilde{Q}_j \gets \mathcal{M}_{\text{local}}\left(\frac{\varepsilon_\tau}{2},\, \text{if } \mathcal{E}(r_j) \geq \varepsilon_\tau \text{ then } r_j \text{ else } \perp\right)$
}

$\widetilde{Q}(D) \gets \mathcal{M}_{\text{curator}}(\{\widetilde{Q}_{ j}\}_{j=1}^n, \, \frac{\beta}{2})$

\Return $\widetilde{Q}(D)$
\end{algorithm}


For utility, we have

\begin{theorem}
\label{the:ldp_fmk_bound}
    For any privacy budget function $\mathcal{E}(\cdot)$, $\beta \in (0,1)$, and any LDP protocol $\mathcal{M}$,  
    Algorithm~\ref{alg:prldp_general} returns an estimate $\widetilde{Q}(D)$ such that, with probability at least $1 - \beta$,
    \begin{equation*}
        |\widetilde{Q}(D) - Q(D)| \leq \underbrace{|Q(D) - Q(D')|}_{\text{(I)}} + \underbrace{\text{\emph{Err}}_{\mathcal{M}}\left(D',\, \varepsilon_{\min}(D)/4,\, \beta/2\right)}_{\text{(II)}},
    \end{equation*}
    where $D'$ is obtained by removing $O\left(\frac{\sqrt{n}}{\varepsilon_{\min}(D)} \log \frac{ \log (\hat{\varepsilon} / \check{\varepsilon}) }{\beta} \right)$
    records from $D$, and $\text{\emph{Err}}_{\mathcal{M}}(\cdot,\, \cdot,\, \cdot)$ denotes the error introduced by the LDP protocol, which depends on the dataset, privacy budget, and failure rate.
\end{theorem}

\begin{proof}
    For privacy, the result follows from the sequential composition of the two components in the framework.  
    For utility, the argument mirrors that of Theorem~\ref{the:fmk_bound}, with the $\sqrt{n}$ term carried over from Theorem~\ref{the:ldpcnt}.  
    In particular, at most $\tilde{O}(\sqrt{n}/\varepsilon_{\min}(D))$ users return $\perp$, and the remaining users contribute through a $\frac{\varepsilon_\tau}{2}$-LDP mechanism.
\end{proof}

The error consists of two components, both of which depend on $\varepsilon_{\min}(D)$. 
Below, we use sum estimation as an example to demonstrate that our mechanism achieves the target error of $O(\text{Err}_{\mathcal{M}}(D,\,$ $\varepsilon_{\min}(D),\,\beta))$.
First, in sum estimation, the SOTA LDP mechanism achieves an error of $\tilde{O}(\sqrt{n} \cdot \text{Max}(D)/\varepsilon_{\min}(D))$~\cite{huang21mean}.
In our mechanism, component I is $\tilde{O}(\sqrt{n} \cdot \text{Max}(D)/\varepsilon_{\min}(D))$, and component II is also $\tilde{O}(\sqrt{n} \cdot \text{Max}(D)/\varepsilon_{\min}(D))$.
This matches exactly the error of the LDP protocol $\mathcal{M}$ in~\cite{huang21mean} on $D$ under a privacy budget of $\varepsilon_{\min}(D)$.

    \section{Experiments}
\label{sec:experiment}

In this section, we evaluate the proposed PrDP algorithms and compare them with existing methods. 
For empirical results of PrLDP, please refer to Appendix~B~\cite{full_version}.
\revthr{We also include experiments with our PrDP framework on an SJA query, two-line path counting, in Appendix~C~\cite{full_version}}

\paragraph{Basic Counting} 
We evaluate our PrDP basic counting algorithm (Algorithm~\ref{alg:cnt}) against two baselines: the naive mechanism using $\check{\varepsilon}$ and the SOTA PDP method presented in~\cite{dajun2024PDP}.
The PDP method is selected for the same reasons discussed earlier in Section \ref{sec:prdp_pdp}. 
For clarification, in this section, the basic counting query aims to privately report the number of records $n$ in the dataset $D$.

\paragraph{Sum Estimation}
For sum estimation, we evaluate our PrDP count (Algorithm~\ref{alg:cnt}) extension and PrDP framework (Algorithm~\ref{alg:gfra}), \revthr{where we use the Laplace-based standard DP sum estimation mechanism from~\cite{dong2023universal} to serve as $\mathcal{M}$.}
We compare them against two baselines: a naive use of $\check{\varepsilon}$ and the SOTA PDP method in~\cite{dajun2024PDP}.

\paragraph{Max} For the max problem, our PrDP framework (Algorithm~\ref{alg:gfra}) is compared with a naive use of $\check{\varepsilon}$, \revthr{where the exponential-based max-selection mechanism from~\cite{asi2020instance,dong2023universal} is used as $\mathcal{M}$.}

\subsection{Setup}
\label{sec:exp_setup}
\paragraph{Datasets}
Both real-world and synthetic datasets are used in this work. 
The synthetic datasets are generated from two distributions: the Normal distribution ($f(x) \propto \exp\left( -((x - \mu)^2) / (2\sigma^2) \right)$) and the Zipf distribution ($f(x) \propto (x + a)^{-b}$).
For the Normal distribution, we set $\mu = 50\text{k},\, \sigma = 50\text{k}$ and $\mu = 500\text{k},\, \sigma = 500\text{k}$. 
For the Zipf distribution, we use $a = 1,\, b = 3$ and $a = 1,\, b = 5$. 
This results in a total of four synthetic bank datasets.
We use four real-world datasets representing practical needs across various financial scenarios, including a real bank dataset~\cite{banking_dataset_2023}, salary payment datasets—San Francisco Salary (SF-Salary)~\cite{sf_salaries_2011_2014} and Ontario Salary (Ont-Salary)~\cite{ontario_dataset_2020}—and a market trade dataset, Japan Trade (JP-Trade)~\cite{million_data_csv_2020}.
\revthr{For clarity and simplicity, all datasets are preprocessed to exclude negative or abnormal values~\footnote{\revthr{Negative values can be supported by shifting the domain by a sufficiently large constant, executing the mechanism over the shifted records, and finally shifting the query results back. 
The privacy budget function must also be defined to accommodate negative inputs—for example, by using absolute values.}}.} 
Since the JP-Trade dataset contains 100 million records, we select only trades to a specific country as a representative subset. 
Table~\ref{table:datasets} presents the statistics of these real-world datasets.
As previously discussed, $U$ is consistently set to $10^{12}$ in all these financial settings, except for the scaling experiments where $U$ varies.
This conservatively large value assumes no prior knowledge of the datasets, thereby ensuring stronger privacy.

\begin{table}[H]
\resizebox{0.55\columnwidth}{!}{
\renewcommand{\arraystretch}{1.5}   

\begin{tabular}{c|c|c|c|c}
\hline
Dataset & $n$                & $\text{Max}(D)$    & mean $D$           & median $D$         \\ \hline
\hline
Bank    & $3.27 \times 10^4$ & $1.02 \times 10^5$ & $1.91 \times 10^3$ & $8.34 \times 10^2$ \\ \hline
SF-Salary  & $1.48 \times 10^5$ & $5.68 \times 10^5$ & $7.52 \times 10^4$ & $7.17 \times 10^4$ \\ \hline
Ont-Salary & $5.75 \times 10^5$ & $1.75 \times 10^6$ & $1.27 \times 10^5$ & $1.16 \times 10^5$ \\ \hline
JP-Trade & $2.03 \times 10^5$ & $ 1.71 \times 10^6$ & $2.20\times 10^3$ & $3.20 \times 10^2$ \\ \hline
\end{tabular}
}
\vspace{0.1in}
\caption{Statistics of real-world datasets.}
\label{table:datasets}
\end{table}

\paragraph{Privacy Specification} 
We evaluate three different types of privacy specifications in total.
For the experiments presented in the tables in this section, we use $\mathcal{E}(r) = 10^4 / v_{\text{Bal}}$ (\textit{inverse}) as the privacy budget function.
For the subsequent figures, where we evaluate performance under varying dataset sizes, $U$, and data distribution, we additionally consider two alternative functions to enrich the privacy specification: 
$\mathcal{E}(r) = 500 / \log^4 v_{\text{Bal}}$ (\textit{log}) 
and 
$\mathcal{E}(r) = 8 / \sqrt{v_{\text{Bal}}}$ (\textit{sqrt}).
We set $\check{\varepsilon} = \mathcal{E}(U)$ and $\hat{\varepsilon} = 100$ consistently across all experiments.

\paragraph{Experimental Parameters} 
All experiments are conducted on a PC equipped with an Apple M4 Max CPU (16 cores) and 128~GB of memory. 
The failure rate is consistently set to $\beta = 10\%$, and each experiment is repeated 50 times. 
We use absolute error for basic counting and sum estimation, and rank error for the max query. 
The average relative error (RE) and running time are reported after discarding the top 20\% and bottom 20\% of the results, thereby conservatively reflecting the $10\%$ failure rate.

\subsection{Results}
\label{sec:exp_results}

\begin{table*}[h]
\resizebox{1.0\columnwidth}{!}{
\renewcommand{\arraystretch}{1.3}  
\begin{tabular}{ccc||cccc||cccc}
\hline
\multicolumn{3}{c||}{\multirow{4}{*}{Dataset}} &
  \multicolumn{4}{c||}{Synthetic Data} &
  \multicolumn{4}{c}{Real-world Data} \\ \cline{4-11} 
\multicolumn{3}{c||}{} &
  \multicolumn{2}{c|}{Normal (200k)} &
  \multicolumn{2}{c||}{Zipf (200k)} &
  \multicolumn{1}{c|}{\multirow{3}{*}{Bank}} &
  \multicolumn{1}{c|}{\multirow{3}{*}{SF-Salary}} &
  \multicolumn{1}{c|}{\multirow{3}{*}{Ont-Salary}} &
  \multirow{3}{*}{JP-Trade} \\ \cline{4-7}
\multicolumn{3}{c||}{} &
  \multicolumn{1}{c|}{$\mu=50$k} &
  \multicolumn{1}{c|}{$\mu=500$k} &
  \multicolumn{1}{c|}{a=1} &
  a=1 &
  \multicolumn{1}{c|}{} &
  \multicolumn{1}{c|}{} &
  \multicolumn{1}{c|}{} &
   \\
\multicolumn{3}{c||}{} &
  \multicolumn{1}{c|}{$\sigma=50$k} &
  \multicolumn{1}{c|}{$\mu=500$k} &
  \multicolumn{1}{c|}{b=3} &
  b=5 &
  \multicolumn{1}{c|}{} &
  \multicolumn{1}{c|}{} &
  \multicolumn{1}{c|}{} &
   \\ \hline
   \hline
\multicolumn{1}{c|}{\multirow{6}{*}{\begin{tabular}[c]{@{}c@{}}Basic \\ Counting\end{tabular}}} &
  \multicolumn{1}{c|}{\multirow{2}{*}{Naive}} &
  RE (\%) &
  \multicolumn{1}{c|}{$>100$} &
  \multicolumn{1}{c|}{$>100$} &
  \multicolumn{1}{c|}{$>100$} &
  $>100$ &
  \multicolumn{1}{c|}{$>100$} &
  \multicolumn{1}{c|}{$>100$} &
  \multicolumn{1}{c|}{$>100$} &
  $>100$ \\
\multicolumn{1}{c|}{} &
  \multicolumn{1}{c|}{} &
  Time (s) &
  \multicolumn{1}{c|}{0.018} &
  \multicolumn{1}{c|}{0.020} &
  \multicolumn{1}{c|}{0.019} &
  0.025 &
  \multicolumn{1}{c|}{0.0017} &
  \multicolumn{1}{c|}{0.013} &
  \multicolumn{1}{c|}{0.035} &
  0.017 \\ \cline{2-11} 
\multicolumn{1}{c|}{} &
  \multicolumn{1}{c|}{\multirow{2}{*}{SOTA PDP}} &
  RE (\%) &
  \multicolumn{1}{c|}{2.27} &
  \multicolumn{1}{c|}{13.69} &
  \multicolumn{1}{c|}{1.28} &
  0.641 &
  \multicolumn{1}{c|}{1.34} &
  \multicolumn{1}{c|}{3.29} &
  \multicolumn{1}{c|}{1.17} &
  0.686 \\
\multicolumn{1}{c|}{} &
  \multicolumn{1}{c|}{} &
  Time (s) &
  \multicolumn{1}{c|}{1.83} &
  \multicolumn{1}{c|}{2.22} &
  \multicolumn{1}{c|}{1.74} &
  1.74 &
  \multicolumn{1}{c|}{0.24} &
  \multicolumn{1}{c|}{1.34} &
  \multicolumn{1}{c|}{5.82} &
  1.25 \\ \cline{2-11} 
\multicolumn{1}{c|}{} &
  \multicolumn{1}{c|}{\multirow{2}{*}{\begin{tabular}[c]{@{}c@{}}PrDP Count\\ (Ours)\end{tabular}}} &
  RE (\%) &
  \multicolumn{1}{c|}{\cellcolor{gray!15}{0.0138}} &
  \multicolumn{1}{c|}{\cellcolor{gray!15}{0.279}} &
  \multicolumn{1}{c|}{\cellcolor{gray!15}{0.00941}} &
  \cellcolor{gray!15}{0.0196} &
  \multicolumn{1}{c|}{\cellcolor{gray!15}{0.0357}} &
  \multicolumn{1}{c|}{\cellcolor{gray!15}{0.0390}} &
  \multicolumn{1}{c|}{\cellcolor{gray!15}{0.0505}} &
  \cellcolor{gray!15}{0.0647} \\
\multicolumn{1}{c|}{} &
  \multicolumn{1}{c|}{} &
  Time (s) &
  \multicolumn{1}{c|}{1.06} &
  \multicolumn{1}{c|}{0.95} &
  \multicolumn{1}{c|}{1.29} &
  1.33 &
  \multicolumn{1}{c|}{0.23} &
  \multicolumn{1}{c|}{0.78} &
  \multicolumn{1}{c|}{2.86} &
  1.41 \\ 
  \hline
  \hline
\multicolumn{1}{c|}{\multirow{8}{*}{\begin{tabular}[c]{@{}c@{}}Sum \\ Estimation\end{tabular}}} &
  \multicolumn{1}{c|}{\multirow{2}{*}{Naive}} &
  RE (\%) &
  \multicolumn{1}{c|}{100} &
  \multicolumn{1}{c|}{100} &
  \multicolumn{1}{c|}{100} &
  100 &
  \multicolumn{1}{c|}{100} &
  \multicolumn{1}{c|}{100} &
  \multicolumn{1}{c|}{100} &
  100 \\
\multicolumn{1}{c|}{} &
  \multicolumn{1}{c|}{} &
  Time (s) &
  \multicolumn{1}{c|}{0.13} &
  \multicolumn{1}{c|}{0.12} &
  \multicolumn{1}{c|}{0.24} &
  0.24 &
  \multicolumn{1}{c|}{0.035} &
  \multicolumn{1}{c|}{0.089} &
  \multicolumn{1}{c|}{0.35} &
  0.12 \\ \cline{2-11} 
\multicolumn{1}{c|}{} &
  \multicolumn{1}{c|}{\multirow{2}{*}{SOTA PDP}} &
  RE (\%) &
  \multicolumn{1}{c|}{10.68} &
  \multicolumn{1}{c|}{48.80} &
  \multicolumn{1}{c|}{17.44} &
   7.18 &
  \multicolumn{1}{c|}{21.25} &
  \multicolumn{1}{c|}{14.92} &
  \multicolumn{1}{c|}{5.40} &
  38.25 \\
\multicolumn{1}{c|}{} &
  \multicolumn{1}{c|}{} &
  Time (s) &
  \multicolumn{1}{c|}{15.17} &
  \multicolumn{1}{c|}{15.25} &
  \multicolumn{1}{c|}{15.12} &
  14.91 &
  \multicolumn{1}{c|}{2.32} &
  \multicolumn{1}{c|}{10.64} &
  \multicolumn{1}{c|}{43.94} &
  14.80 \\ \cline{2-11} 
\multicolumn{1}{c|}{} &
  \multicolumn{1}{c|}{\multirow{2}{*}{\begin{tabular}[c]{@{}c@{}}PrDP Extension\\ (Ours)\end{tabular}}} &
  RE (\%) &
  \multicolumn{1}{c|}{\cellcolor{gray!15}{0.0358}} &
  \multicolumn{1}{c|}{\cellcolor{gray!15}{0.967}} &
  \multicolumn{1}{c|}{\cellcolor{gray!15}{0.198}} &
  \cellcolor{gray!15}{0.321} &
  \multicolumn{1}{c|}{3.54} &
  \multicolumn{1}{c|}{0.984} &
  \multicolumn{1}{c|}{0.244} &
  18.91 \\
\multicolumn{1}{c|}{} &
  \multicolumn{1}{c|}{} &
  Time (s) &
  \multicolumn{1}{c|}{1.13} &
  \multicolumn{1}{c|}{0.98} &
  \multicolumn{1}{c|}{1.40} &
  1.43 &
  \multicolumn{1}{c|}{0.25} &
  \multicolumn{1}{c|}{0.85} &
  \multicolumn{1}{c|}{3.30} &
  1.53 \\ \cline{2-11} 
\multicolumn{1}{c|}{} &
  \multicolumn{1}{c|}{\multirow{2}{*}{\begin{tabular}[c]{@{}c@{}}PrDP Framework\\ (Ours)\end{tabular}}} &
  RE (\%) &
  \multicolumn{1}{c|}{0.187} &
  \multicolumn{1}{c|}{1.65} &
  \multicolumn{1}{c|}{1.04} &
  0.544 &
  \multicolumn{1}{c|}{\cellcolor{gray!15}{2.25}} &
  \multicolumn{1}{c|}{\cellcolor{gray!15}{0.527}} &
  \multicolumn{1}{c|}{\cellcolor{gray!15}{0.225}} &
  \cellcolor{gray!15}{18.23} \\
\multicolumn{1}{c|}{} &
  \multicolumn{1}{c|}{} &
  Time (s) &
  \multicolumn{1}{c|}{1.41} &
  \multicolumn{1}{c|}{1.31} &
  \multicolumn{1}{c|}{1.86} &
  1.95 &
  \multicolumn{1}{c|}{0.33} &
  \multicolumn{1}{c|}{1.06} &
  \multicolumn{1}{c|}{3.98} &
  1.73 \\
  \hline
  \hline
\multicolumn{1}{c|}{\multirow{4}{*}{Max}} &
  \multicolumn{1}{c|}{\multirow{2}{*}{Naive}} &
  RE (\%) &
  \multicolumn{1}{c|}{$>100$} &
  \multicolumn{1}{c|}{$>100$} &
  \multicolumn{1}{c|}{$>100$} &
  $>100$ &
  \multicolumn{1}{c|}{$>100$} &
  \multicolumn{1}{c|}{$>100$} &
  \multicolumn{1}{c|}{$>100$} &
  $>100$ \\
\multicolumn{1}{c|}{} &
  \multicolumn{1}{c|}{} &
  Time (s) &
  \multicolumn{1}{c|}{N.A.} &
  \multicolumn{1}{c|}{N.A.} &
  \multicolumn{1}{c|}{N.A.} &
  N.A. &
  \multicolumn{1}{c|}{N.A.} &
  \multicolumn{1}{c|}{N.A.} &
  \multicolumn{1}{c|}{N.A.} &
  N.A. \\ \cline{2-11} 
\multicolumn{1}{c|}{} &
  \multicolumn{1}{c|}{\multirow{2}{*}{\begin{tabular}[c]{@{}c@{}}PrDP Framework\\ (Ours)\end{tabular}}} &
  RE (\%) &
  \multicolumn{1}{c|}{\cellcolor{gray!15}{0.738}} &
  \multicolumn{1}{c|}{\cellcolor{gray!15}{6.15}} &
  \multicolumn{1}{c|}{\cellcolor{gray!15}{0.734}} &
  \cellcolor{gray!15}{0.387} &
  \multicolumn{1}{c|}{\cellcolor{gray!15}{1.32}} &
  \multicolumn{1}{c|}{\cellcolor{gray!15}{2.02}} &
  \multicolumn{1}{c|}{\cellcolor{gray!15}{0.642}} &
  \cellcolor{gray!15}{0.757} \\
\multicolumn{1}{c|}{} &
  \multicolumn{1}{c|}{} &
  Time (s) &
  \multicolumn{1}{c|}{1.30} &
  \multicolumn{1}{c|}{1.14} &
  \multicolumn{1}{c|}{1.61} &
  1.73 &
  \multicolumn{1}{c|}{0.29} &
  \multicolumn{1}{c|}{0.92} &
  \multicolumn{1}{c|}{3.41} &
  1.60 \\ \hline
\end{tabular}
}
\vspace{0.1in}
\caption{Comparison of different mechanisms under PrDP for basic counting, sum estimation, and max. RE stands for relative error.}
\label{tab:comparison}
\end{table*}

\subsubsection{Utility and Runtime}
Table~\ref{tab:comparison} presents the utility and runtime of our algorithms and corresponding baselines for basic counting, sum estimation, and max queries. 
The results clearly demonstrate that our PrDP mechanisms improve utility across all query types on both synthetic and real-world datasets, while reducing runtime by up to $16\times$. 
For count and sum estimation, our PrDP extension achieves $2\times$ to $50\times$ lower RE, and the PrDP framework achieves $2\times$ to $165\times$ lower RE compared to SOTA PDP methods. 
Despite their relaxed privacy guarantees, our mechanisms provide superior utility.

Regarding our two proposed methods, since we use the \textit{inverse} privacy budget function, both methods theoretically share the same asymptotic error bound. 
However, the PrDP framework allocates part of the privacy budget to estimate $\varepsilon_\tau$, resulting in a higher error than the PrDP extension by a constant factor, as confirmed by our experiments on synthetic datasets.
However, in real-world datasets, we observe that the PrDP framework outperforms the PrDP extension by some small constant factors.
This is because real-world data tends to exhibit more skewed distributions than synthetic datasets, resulting in a greater number of domain partitions with sparse data.
For these sparse partitions, the PrDP framework estimates counts, while the PrDP extension estimates sums. 
Since count estimation generally involves a lower threshold, results are more likely to pass the threshold, leading to fewer discarded records under the PrDP framework.

For the Max problem, the naive use of $\check{\varepsilon}$ 
violates the requirement on $n$ in the base standard DP mechanism of~\cite{asi2020instance, dong2023universal}, which requires $n>\tilde{O}(1/\check{\varepsilon})$ thus rendering the method ineffective. 
This further highlights the superiority of our PrDP framework, which achieves a maximum relative rank error of $6.15\%$ across various datasets while maintaining efficiency.
\vspace{-0.035in}

\subsubsection{Dataset Size and Privacy Budget Functions}

\begin{figure*}[h]
    \centering
    \includegraphics[width=1\textwidth]{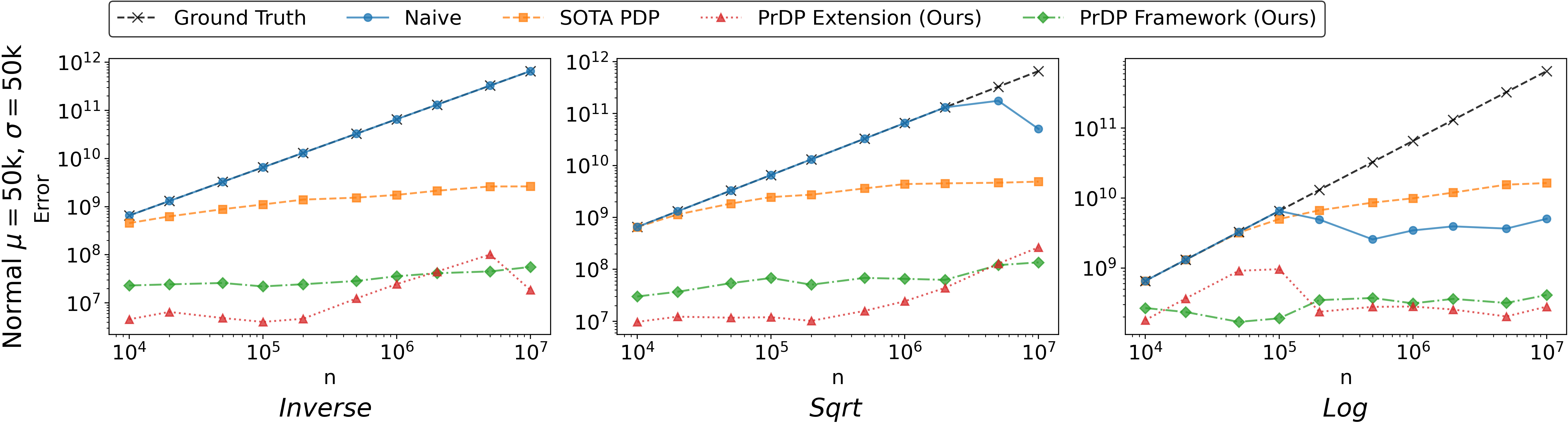}
    \caption{Comparison of errors in sum estimation under PrDP across different mechanisms, with varying dataset sizes $n$. All aixs are in log scale.}
    \label{fig:sum_estimation_n_scaling}
\end{figure*}

\begin{figure*}[h]
    \centering
    \includegraphics[width=1\textwidth]{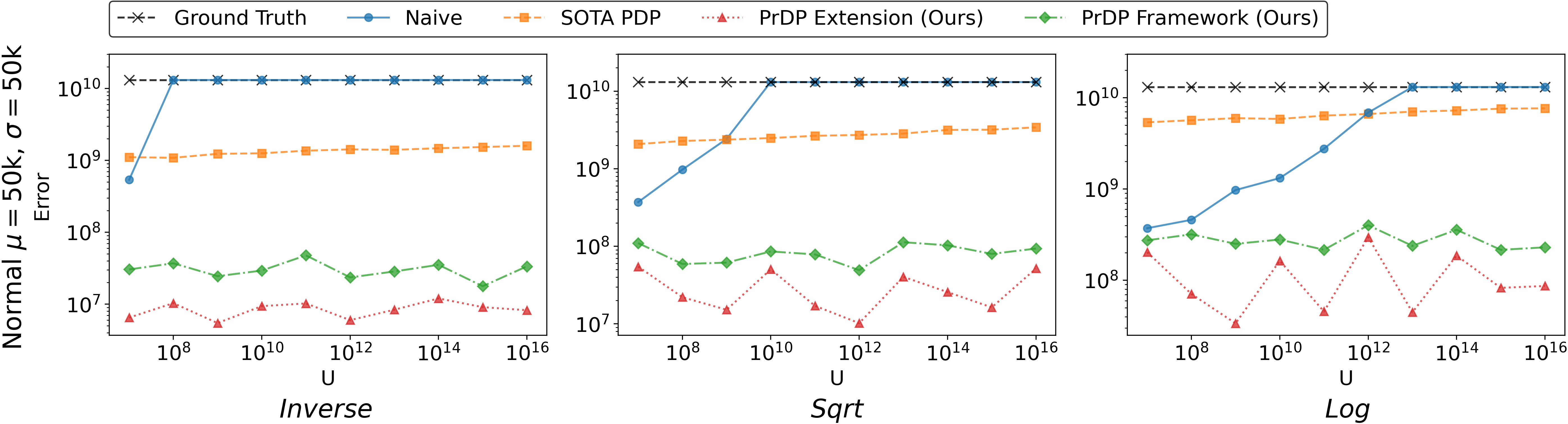}
    \caption{Comparison of errors in sum estimation under PrDP across different mechanisms, with varying $U$. All aixs are in log scale.}
    \label{fig:sum_estimation_u_scaling}
\end{figure*}

\begin{figure*}[h]
    \centering
    \includegraphics[width=1\textwidth]{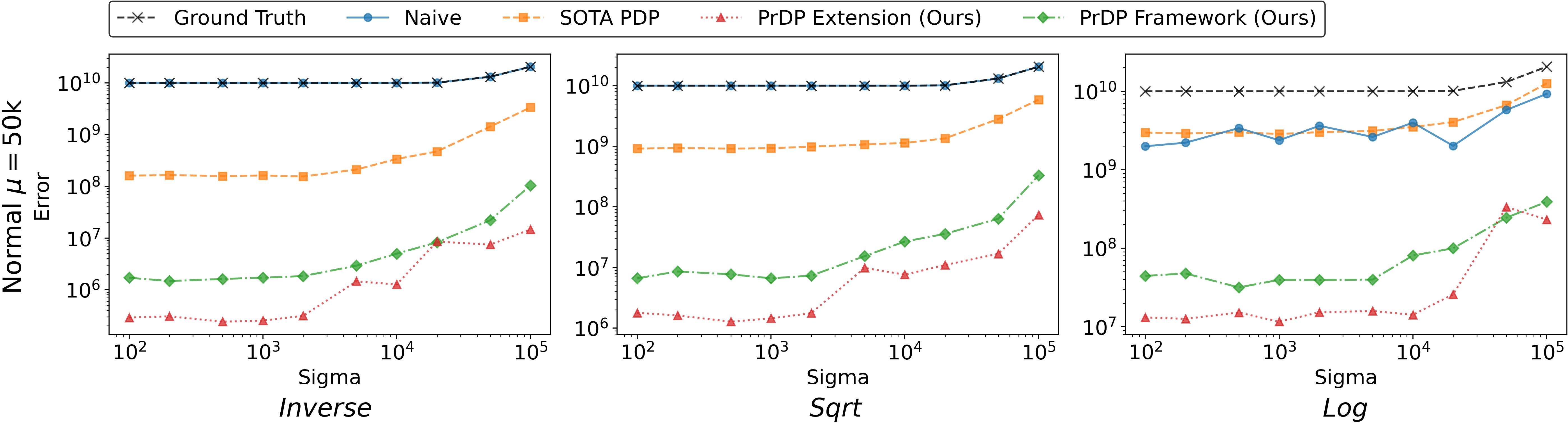}
    \caption{Comparison of errors in sum estimation under PrDP across different mechanisms, with varying $\sigma$. All aixs are in log scale.}
    \label{fig:sum_estimation_sigma_scaling}
\end{figure*}

To evaluate the performance of our algorithms across different dataset sizes and privacy functions, we conduct sum estimation experiments using synthetic datasets drawn from a normal distribution with $\mu=\sigma=50$k, with data sizes from $10^4$ to $10^7$, and apply all three privacy budget functions.  
The results are shown in Figure~\ref{fig:sum_estimation_n_scaling}.
Here, both the PrDP general framework and the extension exhibit high utility across all dataset sizes and outperform the baseline methods. 
Under the \textit{inverse} and \textit{Sqrt} budget function, PrDP extension outperforms the framework by a constant factor in most cases as the framework requires to divide the privacy budget to estimate $\varepsilon_{\tau}$.
However, with \textit{Log} budget functions, the framework outperforms the extension in certain scenarios, as the PrDP extension does not satisfy our target utility guarantee, whereas the framework holds for arbitrary privacy function, as discussed in Section~\ref{sec:ext-sum}.

\vspace{-0.035in}

\subsubsection{Global bound $U$ and Data Skewness}
\label{sec:domain_distribution}
To examine the impact of $U$ and data skewness, we perform two experiments using 200,000-record synthetic datasets. 
In Figure \ref{fig:sum_estimation_u_scaling}, we fix $\mu=\sigma=50$k and vary $U$ from $10^4$ to $10^{16}$. 
The results show that our two proposed methods consistently perform better than the baselines, whereas the baselines exhibit utility degradation as $U$ increases.

In Figure~\ref{fig:sum_estimation_sigma_scaling}, we fix $\mu=50$k and vary $\sigma$ from $10^2$ to $10^5$.
The absolute error increases with data skewness due to two factors: (1) larger $\sigma$ values imply higher $\text{Max}(D)$, necessitating stronger protection (i.e., lower $\varepsilon_{\min}(D)$), and (2) extreme skewness leads to sparse data, making threshold selection more difficult and increasing error.
In all cases, both the PrDP extension and framework consistently outperform baselines.
\revtwo{Note that even under stronger data skewness (larger $\sigma$), the framework outperforms the extension in only a few cases. 
This is because synthetic datasets exhibit smooth tails and therefore cannot replicate the sudden spikes present in real-world data, leading to different results compared to those in Table~\ref{tab:comparison} for real-world datasets.}

    \section{CONCLUSION \& FUTURE WORK}
\label{sec:fu_work}
In this study, we propose the first general framework for per-record privacy requirements in both central (PrDP) and local (PrLDP) settings. 
Our framework accommodates any type of privacy requirement, provided that a corresponding standard DP or LDP mechanism exists. 
This serves as a transparent foundation for addressing per-record privacy needs across a wide range of DP research. 
A current limitation is that the framework supports only static databases. 
An important direction for future work is to extend it to dynamic settings~\cite{dwork2010differential,chan11continual,dong2023continual,dong2024continual,xie2025efficient}, where records with different privacy requirements arrive in a stream and query results are released continuously.
This scenario introduces new challenges in preserving privacy while maintaining utility, as $\varepsilon_{\min}(D)$ evolves over time. 
Another direction is to extend the framework to support the shuffle-DP model~\cite{luo2025rm2,balle2019privacy,cheu2019distributed}, a distributed setting that improves utility by introducing a shuffler between the parties and the analyzer to break the linkage between each record and its corresponding party. Existing work on shuffle-DP has not yet considered the case where the privacy budget is allocated on a per-record basis, which represents a more general and broadly relevant setting.

\section*{Acknowledgments}

This research is supported by the NTU–NAP Startup Grant (024584-00001) and the Singapore Ministry of Education Tier 1 Grant (RG19/25).
We would also like to thank the anonymous reviewers who have made valuable suggestions on improving the presentation of the paper.
}

\clearpage
\bibliographystyle{ACM-Reference-Format}
\bibliography{main_bib} 

\clearpage
\appendix

\section{Details of Sum Estimation Bound for Naive Extension of Algorithm~\ref{alg:cnt} to Other Types of Privacy Budget Functions}  
\label{app:extension}

\begin{theorem}
\label{the:straight_sum}
    The sum estimation algorithm described in Section~\ref{sec:ext-sum} satisfies $\mathcal{E}$-PrDP and it returns $\widetilde{Q}_{\text{\emph{sum}}(D)}$, such that with at least $1 - \beta$ probability, we have
    \begin{equation}
    \label{eqn:general_bound_series}
    \begin{aligned}
        & \quad \quad |\widetilde{Q}_{\text{\emph{sum}}}(D) - \widetilde{Q}_{\text{\emph{sum}}}(D)|  \\ 
         &\leq \sum_{i=1}^{\ell-1} \left|Q_{\text{\emph{sum}}}(D \cap \mathcal{X}_i) \right|
         + \sum_{i=\ell}^{\lceil \log (\hat{\varepsilon} / \check{\varepsilon}) \rceil} 
         \left| \widetilde{Q_{\text{\emph{sum}}}}(D \cap \mathcal{X}_i) - Q_{\text{\emph{sum}}}(D \cap \mathcal{X}_i) \right| \\
         &\leq \sum_{i=k}^{\ell-1} 
         2 \cdot \max_{r \in \mathcal{X}_i} \left( \frac{v_{\text{Bal}}}{\mathcal{E}(r)} \right)  
         \ln\left(\frac{\lceil \log (\hat{\varepsilon} / \check{\varepsilon}) \rceil}{\beta}\right) + \\
         &\sum_{i=\ell}^{\lceil \log (\hat{\varepsilon} / \check{\varepsilon}) \rceil}  
         \max_{r \in \mathcal{X}_i} \left( \frac{v_{\text{Bal}}}{\mathcal{E}(r)} \right)  
         \ln\left(\frac{\lceil \log (\hat{\varepsilon} / \check{\varepsilon}) \rceil}{\beta}\right).
    \end{aligned}
    \end{equation}
    where $k$ is the index of the first non-empty privacy-specified domain.
\end{theorem}
\begin{proof}
    The privacy and error-bound proofs can be derived similarly to those for Theorem~\ref{the:cnt}, with the error bound now modified due to the new threshold.  
\end{proof}

Furthermore, when the privacy budget function is monotonic non-increasing, (\ref{eqn:general_bound_series}) further reduces to
\begin{equation}
\label{eqn:general_bound_first_term}
\begin{aligned}
     (\ref{eqn:general_bound_series}) &  \leq \left( 2 \cdot \frac{\mathcal{E}^{-1}(\varepsilon_{\text{min}}(D)/2)}{\varepsilon_{\text{min}}(D)} + \frac{\mathcal{E}^{-1}(\varepsilon_{\tau})}{\varepsilon_\tau}  \right)
     \ln\left(\frac{\lceil \log (\hat{\varepsilon} / \check{\varepsilon}) \rceil}{\beta}\right) \\
     & \leq O\left(\frac{\mathcal{E}^{-1}(\varepsilon_{\text{min}}(D)/2)}{\varepsilon_{\text{min}}(D)}  
     \log\left(\frac{\log (\hat{\varepsilon} / \check{\varepsilon})}{\beta}\right)\right),
\end{aligned}
\end{equation}
where the inequality in the first line follows from the monotonicity of the privacy budget function, which implies that $\max_{r \in \mathcal{X}_i} \left( \frac{v_{\text{Bal}}}{\mathcal{E}(r)} \right)$ is simply $\frac{\mathcal{E}^{-1}(2^{i-1} \cdot \check{\varepsilon})}{2^{i-1} \cdot \check{\varepsilon}}$.
This monotonicity guarantees that the summation decreases at least exponentially and is therefore dominated solely by its first term.  

We notice that the error bound now depends on the exact form of the privacy budget function.  
Recall that our privacy-specific domain partitioning provides an accurate estimation $\varepsilon_\tau \geq \frac{1}{2} \varepsilon_{\text{min}}(D)$.  
Hence, when $\mathcal{E}(r) = \alpha \cdot v^{-1}$, it follows that $\mathcal{E}^{-1}(\varepsilon_\tau) = \alpha/\varepsilon_\tau \leq \alpha/\frac{1}{2} \varepsilon_{\text{min}}(D) = 2 \text{Max}(D)$.  
Subsequently, the final error is given by  
\begin{equation*}
O\left(\frac{\text{Max}(D)}{\varepsilon_{\text{min}}(D)} \log\left(\frac{\log (\hat{\varepsilon} / \check{\varepsilon})}{\beta}\right)\right)=\tilde{O}\left(\frac{\text{Max}(D)}{\varepsilon_{\text{min}}(D)}\right).
\end{equation*}
So for this specific privacy budget function, we can achieve the target of $\tilde{O}\left(\frac{\text{Max}(D)}{\varepsilon_{\text{min}}(D)}\right)$.

However, when the privacy budget function is defined as $\mathcal{E}(r) = \alpha'/\log (v_{\text{Bal}})$, where $\alpha'$ is a constant parameter,
we have  
\begin{equation}
\begin{aligned}
\label{eqn:log_reverse}
    \mathcal{E}^{-1}\left(\varepsilon_\tau\right) &= \exp\left(\frac{\alpha'}{\varepsilon_\tau}\right) \leq \exp\left(\frac{\alpha'}{\frac{1}{2} \varepsilon_{\text{min}}(D)}\right) \\
    &= \left(\exp\left(\frac{\alpha'}{\varepsilon_{\text{min}}(D)}\right)\right)^2  = \text{Max}(D)^2.
\end{aligned}
\end{equation}  
Substituting (\ref{eqn:log_reverse}) into (\ref{eqn:general_bound_first_term}), we obtain  
\begin{equation*}
O\left(\frac{\text{Max}(D)^2}{\varepsilon_{\text{min}}(D)} \log\left(\frac{\log (\hat{\varepsilon} / \check{\varepsilon})}{\beta}\right)\right) = \tilde{O}\left(\frac{\text{Max}(D)^2}{\varepsilon_{\text{min}}(D)}\right),
\end{equation*}
where a quadratic dependence on $\text{Max}(D)$ emerges, failing to achieve our desired bound.   

The intuition behind the degradation in the error is that we use the inverse of the privacy budget function to calculate the maximum value (and thus the required noise scale) for each domain.
Note for each domain, the left ending point of that domain can differ with the privacy budgets that indeed exist in that domain by a factor of $1/2$. 
The $1/2$ term in Theorem~\ref{the:cnt} becomes negligible when the privacy budget decreases at least inversely proportional to the balance. 
However, when the privacy budget decreases more slowly, say, logarithmically in $v_{\text{Bal}}$, the computed noise scale will deviate by a quadratic factor.

Additionally, in the extreme case where the privacy budget function is either constant (i.e., when the PrDP setting reduces to standard DP) or determined by multiple variables
, the algorithm may group all records into a single privacy-specified domain or mix records with large and small balances within the same domain.  
Consequently, the error bound would depend on $U$, as estimating the minimum privacy budget provides no clear knowledge about the balance.  
This degradation can also be understood from a pure mathematical perspective: many-to-one functions do not have a well-defined inverse function.

\section{PrLDP Basic Counting Experiment}
\label{app:PrLDP_exp}

To evaluate the effectiveness and efficiency of Algorithm~\ref{alg:local_prldp_count} and Algorithm~\ref{alg:analyzer_prldp_count} for basic counting under PrLDP, we conduct experiments under the same settings as in Section~\ref{sec:experiment}, comparing the results to the naive approach using $\check{\varepsilon}$. 
As expected, the baseline provides no utility, whereas our methods efficiently maintain a relative error below 10\% across both synthetic and real-world datasets, despite the error's dependence on $\sqrt{n}$.

\begin{table}[h]
\resizebox{0.6\columnwidth}{!}{
\renewcommand{\arraystretch}{1.1}
\begin{tabular}{ccc||cc||cc}
\hline
\multicolumn{3}{c||}{\multirow{4}{*}{Dataset}} &
  \multicolumn{2}{c||}{Synthetic Data} &
  \multicolumn{2}{c}{Real-world Data} \\ \cline{4-7}
\multicolumn{3}{c||}{} &
  \multicolumn{1}{c|}{Normal (200k)} &
  Zipf (200k) &
  \multicolumn{1}{c|}{\multirow{3}{*}{Bank}} &
  \multirow{3}{*}{JP-Trade} \\ \cline{4-5}
\multicolumn{3}{c||}{} &
  \multicolumn{1}{c|}{$\mu=50$k} &
  $a=1$ &
  \multicolumn{1}{c|}{} &
  \\ 
\multicolumn{3}{c||}{} &
  \multicolumn{1}{c|}{$\sigma=50$k} &
  $b=3$ &
  \multicolumn{1}{c|}{} &
  \\ \hline\hline
\multicolumn{1}{c|}{\multirow{4}{*}{Max}} &
  \multicolumn{1}{c|}{\multirow{2}{*}{Naive}} &
  RE (\%) &
  \multicolumn{1}{c|}{$>100$} &
  $>100$ &
  \multicolumn{1}{c|}{$>100$} &
  $>100$ \\
\multicolumn{1}{c|}{} &
  \multicolumn{1}{c|}{} &
  Time (s) &
  \multicolumn{1}{c|}{0.081} &
  0.077 &
  \multicolumn{1}{c|}{0.12} &
  0.87 \\ \cline{2-7}
\multicolumn{1}{c|}{} &
  \multicolumn{1}{c|}{\multirow{2}{*}{\begin{tabular}[c]{@{}c@{}}PrLDP Count\\ (Ours)\end{tabular}}} &
  RE (\%) &
  \multicolumn{1}{c|}{\cellcolor{gray!15}{9.84}} &
  \cellcolor{gray!15}{8.53} &
  \multicolumn{1}{c|}{\cellcolor{gray!15}{4.51}} &
  \cellcolor{gray!15}{2.84} \\
\multicolumn{1}{c|}{} &
  \multicolumn{1}{c|}{} &
  Time (s) &
  \multicolumn{1}{c|}{7.15} &
  7.31 &
  \multicolumn{1}{c|}{1.15} &
  7.25 \\ \hline
\end{tabular}
}
\vspace{0.1in}
\caption{Performance of PrLDP-Count on real-world datasets.}
\label{tab:prldp_real}
\end{table}

\section{PRDP General Framework Experiments for SJA Queries}
\label{app:sja}
\revthr{
To demonstrate the applicability of our PrDP framework to SJA queries, we select
two-line-path counting as a representative example.
For the base DP mechanism, we uses the \textit{high-order local sensitivity}.}
\revthr{
For the dataset, we build a weighted graph from \textit{email-Eu-core} network~\cite{paranjape2017motifs}, where
each node denotes a research institution and each edge weight is the number of
emails exchanged between the corresponding institutions.
The graph contains roughly 1,000 nodes and more than 15,000 edges.
We define the privacy-budget function as
$\mathcal{E}(\text{edge}) = \tfrac{100}{\text{weight}}$.
The experiment is repeated 50 times.
We discard the top and bottom 10\% of the runs and report the mean RE and the mean runtime.}

\revthr{
Our general framework achieves a mean RE of 2.06\% within 0.10~s.  
Conversely, naive use of $\check{\varepsilon}$ requires substantial noise, pushing the relative error above 100\% in 0.04~s.  
These findings underscore the superior accuracy and efficiency of our approach.}

\end{document}